\documentclass[%
 reprint,superscriptaddress,
 amsmath,amssymb,
 aps]{revtex4-2}

\usepackage{graphicx}
\usepackage{dcolumn}
\usepackage{bm}
\usepackage{siunitx}
\usepackage{braket}
\usepackage{amsmath,amssymb,amsfonts}
\usepackage{hyperref}
\hypersetup{colorlinks=true,allcolors=blue}
\usepackage{todonotes}

\newcommand{\WWU}{Institut f{\"u}r Festk{\"o}rpertheorie, Universit{\"a}t M{\"u}nster, 48149 M{\"u}nster, Germany}
\newcommand{\TUdo}{Condensed Matter Theory, Department of Physics, TU Dortmund, 44221 Dortmund, Germany}
\newcommand{\Bayreuth}{Theoretische Physik III, Universit{\"a}t Bayreuth, 95440 Bayreuth, Germany}
\newcommand{\HWU}{Heriot-Watt
University, Edinburgh EH14 4AS, United Kingdom}
\begin{document}

\preprint{arxiv}


\title{Temperature-independent almost perfect photon entanglement from quantum dots via the SUPER scheme}

\author{Thomas K. Bracht}%
    \affiliation{\WWU}
    \email{t.bracht@wwu.de}
    \affiliation{\TUdo}
\author{Moritz Cygorek}
    \affiliation{\HWU}
\author{Tim Seidelmann}%
    \affiliation{\Bayreuth}
\author{Vollrath Martin Axt}%
    \affiliation{\Bayreuth}
\author{Doris E. Reiter}%
    \affiliation{\TUdo}

\date{\today}

\begin{abstract}
Entangled photon pairs are essential for quantum communication technology. They can be generated on-demand by semiconductor quantum dots, but several mechanisms are known to reduce the degree of entanglement. While some obstacles like the finite fine-structure splitting can be overcome by now, the excitation scheme itself can impair the entanglement fidelity. Here, we demonstrate that the swing-up of quantum emitter population (SUPER) scheme applied to a quantum dot in a cavity yields almost perfectly entangled photons. The entanglement degree remains robust against phonon influences even at elevated temperatures, due to decoupling of the excitation and emission process. With this achievement, quantum dots are ready to be used as entangled photon pair sources in applications requiring high degrees of entanglement up to temperatures of about $\SI{80}{K}$.
\end{abstract}

\maketitle

\section{Introduction}
With their ability to generate entangled photons on-demand \cite{orieux2017semiconductor, stevenson2006semiconductor,Huber2018semiconductor}, quantum dots offer exciting possibilities for advancing the field of quantum communication \cite{vajner2022quantum}. To harness their usefulness for quantum applications, considerable efforts have been dedicated to achieving perfect photon entanglement. The generation process in quantum dots relies on the biexciton-exciton cascade. A major obstacle to obtain perfect entanglement is the fine-structure splitting (FSS) between the quantum dot's single exciton states \cite{hudson2007}. An impaired fidelity can be quantified by the concurrence, which becomes unity only in the ideal case. The issue of the FSS has been successfully addressed by applying external fields \cite{stevenson2006semiconductor,muller2009creating,bennett2010electricfield}, advanced quantum dot growth \cite{hafenbrak2007triggered,muller2014demand}, or via strain tuning \cite{trotta2014highly}. These methods have enabled the generation of entangled photons with a remarkably high concurrence (about $97\,\si{\percent}$ at $\SI{5}{K}$) \cite{huber2018strain}. To achieve perfect entanglement, the preparation process of the biexciton is likewise of paramount importance. The two-photon excitation (TPE) assures the ultrafast, on-demand preparation of the biexciton. However, during the action of the TPE pulse, an optical Stark shift is induced on the exciton levels, which acts as an effective FSS. Accordingly, TPE sets a fundamental limit to the achievable concurrence \cite{seidelmann2022two,basso2023}. This calls for a new scheme to excite the biexciton in an ultrafast way, yet, without affecting the degree of entanglement. 

The recently proposed swing-up of quantum emitter population (SUPER) scheme \cite{bracht21swingup} is a candidate to address the excitation issue.
In the SUPER scheme, off-resonant excitation with two pulses is employed to address the desired state  \cite{bracht21swingup,karli2022super,boos2022coherent}. The off-resonant excitation induces an optical Stark shift of the energies of the target states during the excitation. In Ref.~\cite{heinisch2023arxiv} it was shown that the combination of SUPER with a photonic cavity, as available in different geometries \cite{ota2011spontaneous,reitzenstein2010quantum,lodahl2015interfacing,wang2019ondemand,rickert2023high}, leads to improved photon properties, because emission into the cavity is suppressed during the pulse. The question remains open, if by using SUPER the limitation to the degree of entanglement imposed by TPE can be overcome and perfectly entangled photons can be created.

Another obstacle to overcome for solid-state quantum emitters is the interaction with lattice vibration of the crystal lattice, in particular longitudinal acoustic (LA) phonons \cite{ramsay2010phonon, luker2019review,cosacchi2021accuracy,vannucci2022phonon}. For entangled photons, if the biexciton is initially prepared and there is no FSS, the concurrence is unaffected by LA phonons even at elevated temperatures \cite{carmele,seidelmann2019from}. As soon as this situation is broken, phonons degrade the entanglement, in particular at elevated temperatures \cite{heinze2017polarization,Seidelmann2023,lehner2023beyond}. 

In this paper, we show that entangled photons after excitation of a quantum dot with the SUPER scheme can reach $\SI{99.8}{\percent}$ concurrence even under the influence of phonons. More remarkably, the concurrence of over $\SI{99}{\percent}$ remains for increasing temperatures, up to the temperature of liquid nitrogen at $\SI{77}{K}$. Using SUPER for entangled photon generation is therefore highly promising, even at elevated temperatures, which can for example be employed for satellite-based quantum communication \cite{yin2020entanglement}.

\begin{figure}
    \centering
    \includegraphics[width=0.48\textwidth]{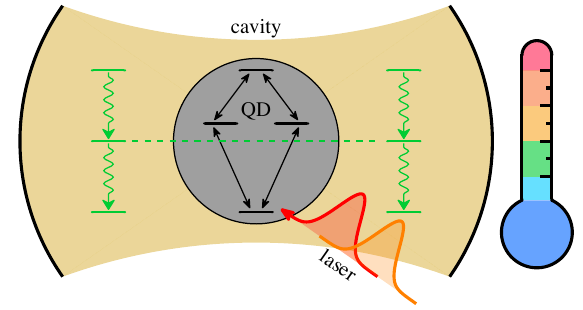}
    \caption{\textbf{Sketch of the system:} A quantum dot (QD) coupled to a two-photon resonant cavity, excited by a diagonally polarized external laser field and subject to electron-phonon interaction. Photons from the QD can either be emitted in free-space directly or to the cavity modes which are subsequently out-coupled. Ideally, the photons coupled out via the cavity are in the maximally entangled state $\ket{\psi} = \frac{1}{\sqrt{2}}(\ket{XX} + \ket{YY})$ which corresponds to the $\ket{\Phi^+}$ Bell-state.}
    \label{fig:fig_cavity}
\end{figure}

\section{Background and model}
Here, we give a brief summary of our model and the simulations. Details of the model and its Hamiltonian alongside the parameters used in the calculations can be found in the appendix (or SI). A sketch of the system is shown in Fig.~\ref{fig:fig_cavity}.
Our model consists of the quantum dot modeled as a four-level system placed inside a photonic cavity. The quantum dot is excited using a diagonally polarized external laser field, treated semi-classically. To maximize the concurrence, our calculations are performed for a quantum dot with zero FSS. The biexciton energy is reduced from twice the single exciton by the biexciton binding energy (BBE), for which we take $\Delta_B=\SI{1}{meV}$ unless stated otherwise. The cavity is set resonant to the two-photon energy, enabling two-photon emission processes \cite{ota2011spontaneous}. We assume a cavity coupling, which for the SUPER scheme yields to high concurrence values as discussed in the appendix. The coupling to LA phonons via the deformation potential coupling, as identified as the main hindering mechanism for state preparation \cite{ramsay2010phonon,luker2019review}, is included via the standard Hamiltonian \cite{reiter2019distinctive}. We further account for radiative decay that does not feed into the cavity via the rate $\gamma$ and cavity losses via the rate $\kappa$. 

To calculate the quantum dot dynamics as well as the dynamics of the cavity photons, we use a process tensor matrix product operator (PT-MPO) method, with details outlined in Ref. \cite{cygorek2022simulation}. Within PT-MPO methods \cite{strathearn2018efficient,jorgensen2019exploiting,pollock2018non,cygorek2022simulation,cygorek2023sublinear} and path integral approaches \cite{Seidelmann2023,cosacchi2018pathintegral} the phonon environment can be included in a numerically complete fashion. Using the PT-MPO method, we can calculate photon properties beyond the limitations inherent to the quantum regression theorem \cite{cosacchi2021accuracy}. We will compare our results to quantum dots without a cavity, where the concurrence is calculated via the quantum dot polarizations \cite{seidelmann2022two}. Calculations without phonons are performed in QuTiP \cite{johansson2012qutip,johansson2013qutip}. From the corresponding dynamics we calculate the correlation functions and the concurrence as detailed in the appendix (or SI).

\section{Concurrence Optimization}

\begin{figure}
    \centering
    \includegraphics{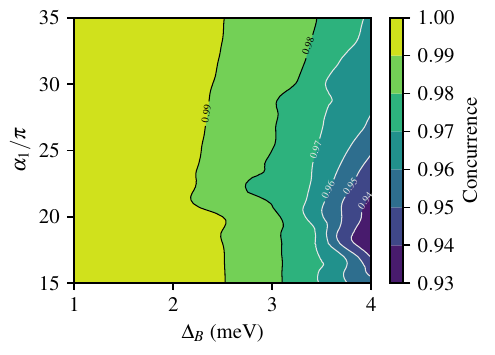}
    \caption{\textbf{Concurrence for the SUPER scheme:}
    Color map of the concurrence as a function of biexciton binding energy $\Delta_B$ and pulse area $\alpha_1$ of the lesser detuned SUPER pulse, for a quantum dot in a cavity, calculated without phonons. The detuning of the lesser detuned pulse is fixed to $\Delta_1=\SI{-5}{meV}$ and the other pulse parameters are optimized numerically towards a high photon yield. For small $\Delta_B$ a concurrence over $\SI{99}{\percent}$ is achieved.
    }
    \label{fig:area1_delta_b}
\end{figure}
We start with the quantum dot in a cavity without phonons. All exciting laser pulses are assumed to be Gaussian with a pulse duration of $\sigma=\SI{2.7}{ps}$ (FWHM of intensity: $\SI{4.5}{ps}$). For TPE without a cavity, this pulse duration results in a concurrence of $\SI{95.1}{\percent}$, in agreement with previous calculations \cite{seidelmann2022two}. Interestingly, for TPE, the cavity does not enhance the concurrence, but only gives a value of $\SI{69.4}{\percent}$. We amount this to the cavity-enhanced photon emission during the pulse, leading to stronger impacts of which-path information and re-excitation.

In SUPER, two pulses with different detunings $\Delta_{1,2}$ with respect to the exciton energy and pulse areas $\alpha_{1,2}$ excite the system. We fix the detuning of the lesser detuned pulse to $\Delta_1=\SI{-5}{meV}$ and then scan the lesser detuned pulse area and numerically search for parameters for the second pulse yielding the highest biexciton occupation. While this does not automatically optimize the concurrence, it ensures that the resulting parameters lead to a high photon yield (cf. appendix). We further consider several BBEs $\Delta_B$, as previous studies revealed the influence of this property \cite{schumacher2012cavity,seidelmann2019phonon}. The results are shown in Fig.~\ref{fig:area1_delta_b}. We find that for $\Delta_B  < \SI{2}{meV}$, the concurrence reaches values above $\SI{99}{\percent}$. A maximum value of $\SI{99.9}{\percent}$ is achieved for every considered $\alpha_1$, when $\Delta_B=\SI{1}{meV}$. Out of these, we choose the parameters that achieve the highest biexciton preparation fidelity, which is at $\alpha_1=32\pi$ and for the higher detuned pulse $\Delta_2=12.96$~meV and $\alpha_2=12.8~\pi$, (cf. also Tab.~\ref{tab:parameters} in appendix). Without a cavity, these parameters give a concurrence of only $\SI{93.1}{\percent}$ due to the induced which-path information during the pulse \cite{seidelmann2022two}.

\begin{figure}
    \centering
    \includegraphics{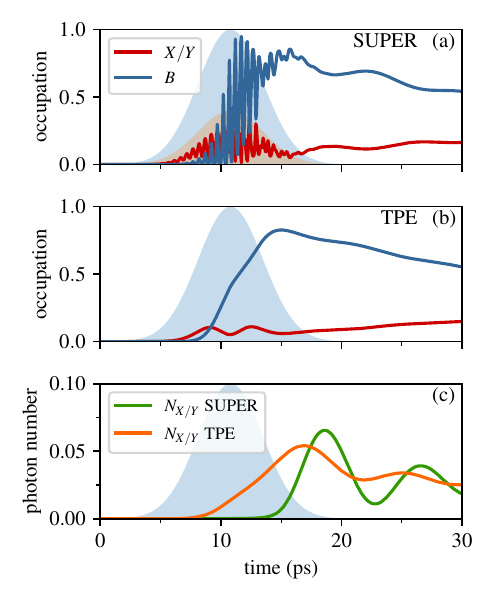}
    \caption{\textbf{Dynamics of the excitation}: For (a) SUPER scheme and for (b) TPE we show the occupations $X$ and $Y$ of the exciton states and $B$ of the biexciton state. (c) Photon numbers $N_{X/Y}$ in the two cavity modes. The shaded background indicates the exciting pulses. In SUPER the photon occupation only rises after the pulses, while in TPE the photon occupation rises already during the pulse.}
    \label{fig:dynamics_compare_cavity}
\end{figure}

The close-to-unity concurrence for SUPER can be traced back to the decoupling of the emission during the pulse due to the Stark shifts of the biexciton-ground state transition. Since the cavity is decoupled during the excitation process, the emission sets in only after the preparation is completed. For the emitted photons, the situation comes close to an initial value problem, where the biexciton is assumed to be initially populated, disregarding the excitation process. This boosts the concurrence because in the initial value problem, the situation is symmetric without any which-path information. In fact, for a true initial value problem without FSS, the concurrence is known to be exactly one \cite{carmele,seidelmann2019from}. \\
This decoupling is visualized in Fig.~\ref{fig:dynamics_compare_cavity}, where the dynamics of the quantum dot states and the cavity photon number are shown. For SUPER, shown in Fig.~\ref{fig:dynamics_compare_cavity}(a), the quantum dot population exhibits the typical swing-up behavior, initially transitioning to the exciton states $X/Y$ before progressing to the biexciton state. For TPE, shown in Fig.~\ref{fig:dynamics_compare_cavity}(b), a monotonic rise of the biexciton occupation is found, while there is also a transient occupation of the exciton states. Due to the diagonal polarization and the vanishing FSS, the $X$ and $Y$ excitons (and also cavity photons) are always addressed equally. The population of the biexciton decays exponentially, accompanied by an additional oscillation of the occupation resulting from the QD-cavity coupling.

A crucial difference of SUPER and TPE lies in the number of cavity photons $N_{X/Y}$ during the preparation process as displayed in Fig~\ref{fig:dynamics_compare_cavity}(c). For SUPER, due to the decoupling there is minimal photon emission into the cavity during the pulses up to approximately $t=\SI{15}{ps}$. After that, we see that the photon number rises, as the cavity and the ground-to-biexciton two-photon transition are resonant again. On the other hand, for TPE, where the cavity is resonant to the relevant transition all the time, the photon numbers $N_{X/Y}$ already rise strongly during the pulse, resulting in the reduced concurrence \cite{seidelmann2022two}. This decoupling effect results in the possibility to achieve nearly perfect entanglement of photons generated from a quantum dot. 

\section{Temperature dependence of the Concurrence}
\begin{figure}[t]
    \centering
    \includegraphics{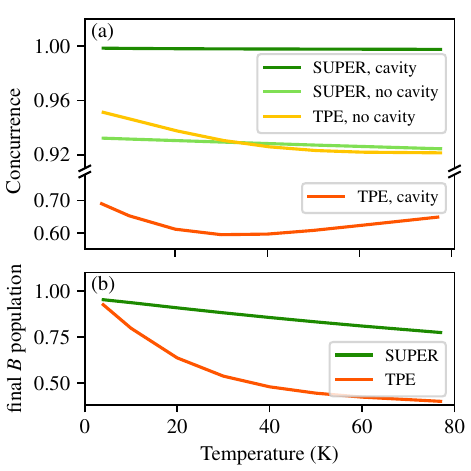}\\
    \caption{\textbf{Temperature dependence of the concurrence}: (a) Concurrence as function of temperature for SUPER and TPE, for the quantum dot either in a cavity or without a cavity.
    (b) Final population of the biexciton state for the SUPER scheme and TPE, with the same parameters as in Fig.~\ref{fig:dynamics_compare_cavity}. To compare the final population values, all decay mechanisms (radiative decay, coupling to the cavity) are neglected.
    }
    \label{fig:temperature_dependence}
\end{figure}

To use quantum dots for practical applications, it is desirable that they work at elevated temperatures. However, with increasing temperature, phonon effects become more pronounced for optical excitation schemes, unless one works in the reappearance regime \cite{vagov2007nonmonotonic,reiter2019distinctive,kaldewey2017demonstrating,vannucci2022phonon}. Phonon coupling can reduce the preparation fidelity of the targeted state drastically \cite{ramsay2010phonon,luker2019review}. As shown in Fig.~\ref{fig:temperature_dependence}(b), the final biexciton occupation decreases as a function of temperature. It is evident that the excitation using the SUPER scheme is less prone to disturbance by phonon interaction. For SUPER, the biexciton population drops approximately linearly with rising temperature, while for TPE, the population rapidly drops below $\SI{50}{\percent}$.

Let us now turn to the concurrence. Interestingly, it has been shown that in the case of zero FSS and an initially prepared biexciton, due to the highly symmetric situation, LA phonons with pure-dephasing type coupling do not affect the concurrence \cite{carmele, seidelmann2019from}. But as soon as the excitation induces an asymmetry in the exciton energies via the Stark shift, phonons degrade the entanglement even further \cite{Seidelmann2023}. Considering the case without cavity, we clearly see in Fig.~\ref{fig:temperature_dependence}(a) that with increasing temperature the concurrence drops as a function of temperature. Like for the populations, the concurrence drops more rapidly for TPE in comparison to SUPER. Including a cavity, TPE exhibits a decline in the concurrence as the temperature increases, in agreement with findings from previous studies \cite{heinze2017polarization,seidelmann2019from} that identified phonons as being a substantial source of decoherence, leading to a reduction of the concurrence. It was also found in Ref.~\cite{seidelmann2019phonon}, that phonons cause a renormalization of the dot-cavity coupling that can improve the concurrence. We attribute the slight increase of the concurrence at around $\SI{50}{K}$ to these effects.

The case is quite different when applying the SUPER scheme on the quantum dot in a cavity. Here, the state preparation and the photon emission are decoupled, hence, the entanglement properties should be similar to the case of an initially prepared biexciton \cite{carmele, seidelmann2019from}.
Remarkably, the concurrence remains at $C>\SI{99.7}{\percent}$ independent of the temperature. This outcome is significant, as it suggests that using this scheme, near-perfect entanglement can be achieved even at elevated temperatures. Our model focussing on the coupling to longitudinal acoustic phonons should be valid to describe the physics up to temperatures of about $\SI{80}{K}$. For higher temperatures, multi longitudinal optical phonon couplings of discrete dot states to the continuum of wetting layer states have been shown to limit the concurrence even for initial value problems and vanishing FSS \cite{carmele2010formation}. We also consider strongly confined quantum dots, because for weakly confined quantum dots, hot exciton states can decrease the concurrence \cite{lehner2023beyond}. 

\section{Conclusions}
We have shown that exciting a semiconductor quantum dot in a cavity via the SUPER scheme overcomes a significant hurdle in generating perfectly entangled photons, namely the limit of the concurrence induced by the duration of the TPE excitation scheme. Our scheme delivers unprecedentedly high values of the concurrence. Strikingly, the almost perfect entanglement can be achieved over a broad parameter range and up to elevated temperatures which paves the way to new types of applications.

\section{Acknowledgements}
We acknowledge financial support from the German Research Foundation DFG through project 428026575 (AEQuDot). 

\newpage

\bibliography{bibfile}

\begin{thebibliography}{52}%
\makeatletter
\providecommand \@ifxundefined [1]{%
 \@ifx{#1\undefined}
}%
\providecommand \@ifnum [1]{%
 \ifnum #1\expandafter \@firstoftwo
 \else \expandafter \@secondoftwo
 \fi
}%
\providecommand \@ifx [1]{%
 \ifx #1\expandafter \@firstoftwo
 \else \expandafter \@secondoftwo
 \fi
}%
\providecommand \natexlab [1]{#1}%
\providecommand \enquote  [1]{``#1''}%
\providecommand \bibnamefont  [1]{#1}%
\providecommand \bibfnamefont [1]{#1}%
\providecommand \citenamefont [1]{#1}%
\providecommand \href@noop [0]{\@secondoftwo}%
\providecommand \href [0]{\begingroup \@sanitize@url \@href}%
\providecommand \@href[1]{\@@startlink{#1}\@@href}%
\providecommand \@@href[1]{\endgroup#1\@@endlink}%
\providecommand \@sanitize@url [0]{\catcode `\\12\catcode `\$12\catcode
  `\&12\catcode `\#12\catcode `\^12\catcode `\_12\catcode `\%12\relax}%
\providecommand \@@startlink[1]{}%
\providecommand \@@endlink[0]{}%
\providecommand \url  [0]{\begingroup\@sanitize@url \@url }%
\providecommand \@url [1]{\endgroup\@href {#1}{\urlprefix }}%
\providecommand \urlprefix  [0]{URL }%
\providecommand \Eprint [0]{\href }%
\providecommand \doibase [0]{https://doi.org/}%
\providecommand \selectlanguage [0]{\@gobble}%
\providecommand \bibinfo  [0]{\@secondoftwo}%
\providecommand \bibfield  [0]{\@secondoftwo}%
\providecommand \translation [1]{[#1]}%
\providecommand \BibitemOpen [0]{}%
\providecommand \bibitemStop [0]{}%
\providecommand \bibitemNoStop [0]{.\EOS\space}%
\providecommand \EOS [0]{\spacefactor3000\relax}%
\providecommand \BibitemShut  [1]{\csname bibitem#1\endcsname}%
\let\auto@bib@innerbib\@empty
\bibitem [{\citenamefont {Orieux}\ \emph {et~al.}(2017)\citenamefont {Orieux},
  \citenamefont {Versteegh}, \citenamefont {Jöns},\ and\ \citenamefont
  {Ducci}}]{orieux2017semiconductor}%
  \BibitemOpen
  \bibfield  {author} {\bibinfo {author} {\bibfnamefont {A.}~\bibnamefont
  {Orieux}}, \bibinfo {author} {\bibfnamefont {M.~A.~M.}\ \bibnamefont
  {Versteegh}}, \bibinfo {author} {\bibfnamefont {K.~D.}\ \bibnamefont
  {Jöns}},\ and\ \bibinfo {author} {\bibfnamefont {S.}~\bibnamefont {Ducci}},\
  }\bibfield  {title} {\bibinfo {title} {Semiconductor devices for entangled
  photon pair generation: a review},\ }\href
  {https://doi.org/10.1088/1361-6633/aa6955} {\bibfield  {journal} {\bibinfo
  {journal} {Reports on Progress in Physics}\ }\textbf {\bibinfo {volume}
  {80}},\ \bibinfo {pages} {076001} (\bibinfo {year} {2017})}\BibitemShut
  {NoStop}%
\bibitem [{\citenamefont {Stevenson}\ \emph {et~al.}(2006)\citenamefont
  {Stevenson}, \citenamefont {Young}, \citenamefont {Atkinson}, \citenamefont
  {Cooper}, \citenamefont {Ritchie},\ and\ \citenamefont
  {Shields}}]{stevenson2006semiconductor}%
  \BibitemOpen
  \bibfield  {author} {\bibinfo {author} {\bibfnamefont {R.~M.}\ \bibnamefont
  {Stevenson}}, \bibinfo {author} {\bibfnamefont {R.~J.}\ \bibnamefont
  {Young}}, \bibinfo {author} {\bibfnamefont {P.}~\bibnamefont {Atkinson}},
  \bibinfo {author} {\bibfnamefont {K.}~\bibnamefont {Cooper}}, \bibinfo
  {author} {\bibfnamefont {D.~A.}\ \bibnamefont {Ritchie}},\ and\ \bibinfo
  {author} {\bibfnamefont {A.~J.}\ \bibnamefont {Shields}},\ }\bibfield
  {title} {\bibinfo {title} {A semiconductor source of triggered entangled
  photon pairs},\ }\href {https://doi.org/10.1038/nature04446} {\bibfield
  {journal} {\bibinfo  {journal} {Nature}\ }\textbf {\bibinfo {volume} {439}},\
  \bibinfo {pages} {179} (\bibinfo {year} {2006})}\BibitemShut {NoStop}%
\bibitem [{\citenamefont {Huber}\ \emph
  {et~al.}(2018{\natexlab{a}})\citenamefont {Huber}, \citenamefont {Reindl},
  \citenamefont {Aberl}, \citenamefont {Rastelli},\ and\ \citenamefont
  {Trotta}}]{Huber2018semiconductor}%
  \BibitemOpen
  \bibfield  {author} {\bibinfo {author} {\bibfnamefont {D.}~\bibnamefont
  {Huber}}, \bibinfo {author} {\bibfnamefont {M.}~\bibnamefont {Reindl}},
  \bibinfo {author} {\bibfnamefont {J.}~\bibnamefont {Aberl}}, \bibinfo
  {author} {\bibfnamefont {A.}~\bibnamefont {Rastelli}},\ and\ \bibinfo
  {author} {\bibfnamefont {R.}~\bibnamefont {Trotta}},\ }\bibfield  {title}
  {\bibinfo {title} {Semiconductor quantum dots as an ideal source of
  polarization-entangled photon pairs on-demand: a review},\ }\href
  {https://doi.org/10.1088/2040-8986/aac4c4} {\bibfield  {journal} {\bibinfo
  {journal} {J. Opt.}\ }\textbf {\bibinfo {volume} {20}},\ \bibinfo {pages}
  {073002} (\bibinfo {year} {2018}{\natexlab{a}})}\BibitemShut {NoStop}%
\bibitem [{\citenamefont {Vajner}\ \emph {et~al.}(2022)\citenamefont {Vajner},
  \citenamefont {Rickert}, \citenamefont {Gao}, \citenamefont {Kaymazlar},\
  and\ \citenamefont {Heindel}}]{vajner2022quantum}%
  \BibitemOpen
  \bibfield  {author} {\bibinfo {author} {\bibfnamefont {D.~A.}\ \bibnamefont
  {Vajner}}, \bibinfo {author} {\bibfnamefont {L.}~\bibnamefont {Rickert}},
  \bibinfo {author} {\bibfnamefont {T.}~\bibnamefont {Gao}}, \bibinfo {author}
  {\bibfnamefont {K.}~\bibnamefont {Kaymazlar}},\ and\ \bibinfo {author}
  {\bibfnamefont {T.}~\bibnamefont {Heindel}},\ }\bibfield  {title} {\bibinfo
  {title} {Quantum communication using semiconductor quantum dots},\ }\href
  {https://doi.org/10.1002/qute.202100116} {\bibfield  {journal} {\bibinfo
  {journal} {Adv. Quantum Technol.}\ ,\ \bibinfo {pages} {2100116}} (\bibinfo
  {year} {2022})}\BibitemShut {NoStop}%
\bibitem [{\citenamefont {Hudson}\ \emph {et~al.}(2007)\citenamefont {Hudson},
  \citenamefont {Stevenson}, \citenamefont {Bennett}, \citenamefont {Young},
  \citenamefont {Nicoll}, \citenamefont {Atkinson}, \citenamefont {Cooper},
  \citenamefont {Ritchie},\ and\ \citenamefont {Shields}}]{hudson2007}%
  \BibitemOpen
  \bibfield  {author} {\bibinfo {author} {\bibfnamefont {A.~J.}\ \bibnamefont
  {Hudson}}, \bibinfo {author} {\bibfnamefont {R.~M.}\ \bibnamefont
  {Stevenson}}, \bibinfo {author} {\bibfnamefont {A.~J.}\ \bibnamefont
  {Bennett}}, \bibinfo {author} {\bibfnamefont {R.~J.}\ \bibnamefont {Young}},
  \bibinfo {author} {\bibfnamefont {C.~A.}\ \bibnamefont {Nicoll}}, \bibinfo
  {author} {\bibfnamefont {P.}~\bibnamefont {Atkinson}}, \bibinfo {author}
  {\bibfnamefont {K.}~\bibnamefont {Cooper}}, \bibinfo {author} {\bibfnamefont
  {D.~A.}\ \bibnamefont {Ritchie}},\ and\ \bibinfo {author} {\bibfnamefont
  {A.~J.}\ \bibnamefont {Shields}},\ }\bibfield  {title} {\bibinfo {title}
  {Coherence of an entangled exciton-photon state},\ }\href
  {https://doi.org/10.1103/PhysRevLett.99.266802} {\bibfield  {journal}
  {\bibinfo  {journal} {Phys. Rev. Lett.}\ }\textbf {\bibinfo {volume} {99}},\
  \bibinfo {pages} {266802} (\bibinfo {year} {2007})}\BibitemShut {NoStop}%
\bibitem [{\citenamefont {Muller}\ \emph {et~al.}(2009)\citenamefont {Muller},
  \citenamefont {Fang}, \citenamefont {Lawall},\ and\ \citenamefont
  {Solomon}}]{muller2009creating}%
  \BibitemOpen
  \bibfield  {author} {\bibinfo {author} {\bibfnamefont {A.}~\bibnamefont
  {Muller}}, \bibinfo {author} {\bibfnamefont {W.}~\bibnamefont {Fang}},
  \bibinfo {author} {\bibfnamefont {J.}~\bibnamefont {Lawall}},\ and\ \bibinfo
  {author} {\bibfnamefont {G.~S.}\ \bibnamefont {Solomon}},\ }\bibfield
  {title} {\bibinfo {title} {Creating polarization-entangled photon pairs from
  a semiconductor quantum dot using the optical stark effect},\ }\href
  {https://doi.org/10.1103/PhysRevLett.103.217402} {\bibfield  {journal}
  {\bibinfo  {journal} {Phys. Rev. Lett.}\ }\textbf {\bibinfo {volume} {103}},\
  \bibinfo {pages} {217402} (\bibinfo {year} {2009})}\BibitemShut {NoStop}%
\bibitem [{\citenamefont {Bennett}\ \emph {et~al.}(2010)\citenamefont
  {Bennett}, \citenamefont {Pooley}, \citenamefont {Stevenson}, \citenamefont
  {Ward}, \citenamefont {Patel}, \citenamefont {de~la Giroday}, \citenamefont
  {Sk{\"o}ld}, \citenamefont {Farrer}, \citenamefont {Nicoll}, \citenamefont
  {Ritchie},\ and\ \citenamefont {Shields}}]{bennett2010electricfield}%
  \BibitemOpen
  \bibfield  {author} {\bibinfo {author} {\bibfnamefont {A.~J.}\ \bibnamefont
  {Bennett}}, \bibinfo {author} {\bibfnamefont {M.~A.}\ \bibnamefont {Pooley}},
  \bibinfo {author} {\bibfnamefont {R.~M.}\ \bibnamefont {Stevenson}}, \bibinfo
  {author} {\bibfnamefont {M.~B.}\ \bibnamefont {Ward}}, \bibinfo {author}
  {\bibfnamefont {R.~B.}\ \bibnamefont {Patel}}, \bibinfo {author}
  {\bibfnamefont {A.~B.}\ \bibnamefont {de~la Giroday}}, \bibinfo {author}
  {\bibfnamefont {N.}~\bibnamefont {Sk{\"o}ld}}, \bibinfo {author}
  {\bibfnamefont {I.}~\bibnamefont {Farrer}}, \bibinfo {author} {\bibfnamefont
  {C.~A.}\ \bibnamefont {Nicoll}}, \bibinfo {author} {\bibfnamefont {D.~A.}\
  \bibnamefont {Ritchie}},\ and\ \bibinfo {author} {\bibfnamefont {A.~J.}\
  \bibnamefont {Shields}},\ }\bibfield  {title} {\bibinfo {title}
  {Electric-field-induced coherent coupling of the exciton states in a single
  quantum dot},\ }\href {https://doi.org/10.1038/nphys1780} {\bibfield
  {journal} {\bibinfo  {journal} {Nature Physics}\ }\textbf {\bibinfo {volume}
  {6}},\ \bibinfo {pages} {947} (\bibinfo {year} {2010})}\BibitemShut {NoStop}%
\bibitem [{\citenamefont {Hafenbrak}\ \emph {et~al.}(2007)\citenamefont
  {Hafenbrak}, \citenamefont {Ulrich}, \citenamefont {Michler}, \citenamefont
  {Wang}, \citenamefont {Rastelli},\ and\ \citenamefont
  {Schmidt}}]{hafenbrak2007triggered}%
  \BibitemOpen
  \bibfield  {author} {\bibinfo {author} {\bibfnamefont {R.}~\bibnamefont
  {Hafenbrak}}, \bibinfo {author} {\bibfnamefont {S.~M.}\ \bibnamefont
  {Ulrich}}, \bibinfo {author} {\bibfnamefont {P.}~\bibnamefont {Michler}},
  \bibinfo {author} {\bibfnamefont {L.}~\bibnamefont {Wang}}, \bibinfo {author}
  {\bibfnamefont {A.}~\bibnamefont {Rastelli}},\ and\ \bibinfo {author}
  {\bibfnamefont {O.~G.}\ \bibnamefont {Schmidt}},\ }\bibfield  {title}
  {\bibinfo {title} {Triggered polarization-entangled photon pairs from a
  single quantum dot up to 30{\hspace{0.167em}}{K}},\ }\href
  {https://doi.org/10.1088/1367-2630/9/9/315} {\bibfield  {journal} {\bibinfo
  {journal} {New J. Phys.}\ }\textbf {\bibinfo {volume} {9}},\ \bibinfo {pages}
  {315} (\bibinfo {year} {2007})}\BibitemShut {NoStop}%
\bibitem [{\citenamefont {M{\"u}ller}\ \emph {et~al.}(2014)\citenamefont
  {M{\"u}ller}, \citenamefont {Bounouar}, \citenamefont {J{\"o}ns},
  \citenamefont {Gl{\"a}ssl},\ and\ \citenamefont
  {Michler}}]{muller2014demand}%
  \BibitemOpen
  \bibfield  {author} {\bibinfo {author} {\bibfnamefont {M.}~\bibnamefont
  {M{\"u}ller}}, \bibinfo {author} {\bibfnamefont {S.}~\bibnamefont
  {Bounouar}}, \bibinfo {author} {\bibfnamefont {K.~D.}\ \bibnamefont
  {J{\"o}ns}}, \bibinfo {author} {\bibfnamefont {M.}~\bibnamefont
  {Gl{\"a}ssl}},\ and\ \bibinfo {author} {\bibfnamefont {P.}~\bibnamefont
  {Michler}},\ }\bibfield  {title} {\bibinfo {title} {On-demand generation of
  indistinguishable polarization-entangled photon pairs},\ }\href
  {https://doi.org/10.1038/nphoton.2013.377} {\bibfield  {journal} {\bibinfo
  {journal} {Nat. Photonics}\ }\textbf {\bibinfo {volume} {8}},\ \bibinfo
  {pages} {224} (\bibinfo {year} {2014})}\BibitemShut {NoStop}%
\bibitem [{\citenamefont {Trotta}\ \emph {et~al.}(2014)\citenamefont {Trotta},
  \citenamefont {Wildmann}, \citenamefont {Zallo}, \citenamefont {Schmidt},\
  and\ \citenamefont {Rastelli}}]{trotta2014highly}%
  \BibitemOpen
  \bibfield  {author} {\bibinfo {author} {\bibfnamefont {R.}~\bibnamefont
  {Trotta}}, \bibinfo {author} {\bibfnamefont {J.~S.}\ \bibnamefont
  {Wildmann}}, \bibinfo {author} {\bibfnamefont {E.}~\bibnamefont {Zallo}},
  \bibinfo {author} {\bibfnamefont {O.~G.}\ \bibnamefont {Schmidt}},\ and\
  \bibinfo {author} {\bibfnamefont {A.}~\bibnamefont {Rastelli}},\ }\bibfield
  {title} {\bibinfo {title} {Highly entangled photons from hybrid
  piezoelectric-semiconductor quantum dot devices},\ }\href
  {https://doi.org/10.1021/nl500968k} {\bibfield  {journal} {\bibinfo
  {journal} {Nano Lett.}\ }\textbf {\bibinfo {volume} {14}},\ \bibinfo {pages}
  {3439} (\bibinfo {year} {2014})}\BibitemShut {NoStop}%
\bibitem [{\citenamefont {Huber}\ \emph
  {et~al.}(2018{\natexlab{b}})\citenamefont {Huber}, \citenamefont {Reindl},
  \citenamefont {Covre~da Silva}, \citenamefont {Schimpf}, \citenamefont
  {Mart\'{\i}n-S\'anchez}, \citenamefont {Huang}, \citenamefont {Piredda},
  \citenamefont {Edlinger}, \citenamefont {Rastelli},\ and\ \citenamefont
  {Trotta}}]{huber2018strain}%
  \BibitemOpen
  \bibfield  {author} {\bibinfo {author} {\bibfnamefont {D.}~\bibnamefont
  {Huber}}, \bibinfo {author} {\bibfnamefont {M.}~\bibnamefont {Reindl}},
  \bibinfo {author} {\bibfnamefont {S.~F.}\ \bibnamefont {Covre~da Silva}},
  \bibinfo {author} {\bibfnamefont {C.}~\bibnamefont {Schimpf}}, \bibinfo
  {author} {\bibfnamefont {J.}~\bibnamefont {Mart\'{\i}n-S\'anchez}}, \bibinfo
  {author} {\bibfnamefont {H.}~\bibnamefont {Huang}}, \bibinfo {author}
  {\bibfnamefont {G.}~\bibnamefont {Piredda}}, \bibinfo {author} {\bibfnamefont
  {J.}~\bibnamefont {Edlinger}}, \bibinfo {author} {\bibfnamefont
  {A.}~\bibnamefont {Rastelli}},\ and\ \bibinfo {author} {\bibfnamefont
  {R.}~\bibnamefont {Trotta}},\ }\bibfield  {title} {\bibinfo {title}
  {Strain-tunable gaas quantum dot: A nearly dephasing-free source of entangled
  photon pairs on demand},\ }\href
  {https://doi.org/10.1103/PhysRevLett.121.033902} {\bibfield  {journal}
  {\bibinfo  {journal} {Phys. Rev. Lett.}\ }\textbf {\bibinfo {volume} {121}},\
  \bibinfo {pages} {033902} (\bibinfo {year} {2018}{\natexlab{b}})}\BibitemShut
  {NoStop}%
\bibitem [{\citenamefont {Seidelmann}\ \emph {et~al.}(2022)\citenamefont
  {Seidelmann}, \citenamefont {Schimpf}, \citenamefont {Bracht}, \citenamefont
  {Cosacchi}, \citenamefont {Vagov}, \citenamefont {Rastelli}, \citenamefont
  {Reiter},\ and\ \citenamefont {Axt}}]{seidelmann2022two}%
  \BibitemOpen
  \bibfield  {author} {\bibinfo {author} {\bibfnamefont {T.}~\bibnamefont
  {Seidelmann}}, \bibinfo {author} {\bibfnamefont {C.}~\bibnamefont {Schimpf}},
  \bibinfo {author} {\bibfnamefont {T.~K.}\ \bibnamefont {Bracht}}, \bibinfo
  {author} {\bibfnamefont {M.}~\bibnamefont {Cosacchi}}, \bibinfo {author}
  {\bibfnamefont {A.}~\bibnamefont {Vagov}}, \bibinfo {author} {\bibfnamefont
  {A.}~\bibnamefont {Rastelli}}, \bibinfo {author} {\bibfnamefont {D.~E.}\
  \bibnamefont {Reiter}},\ and\ \bibinfo {author} {\bibfnamefont {V.~M.}\
  \bibnamefont {Axt}},\ }\bibfield  {title} {\bibinfo {title} {Two-photon
  excitation sets limit to entangled photon pair generation from quantum
  emitters},\ }\href {https://doi.org/10.1103/PhysRevLett.129.193604}
  {\bibfield  {journal} {\bibinfo  {journal} {Phys. Rev. Lett.}\ }\textbf
  {\bibinfo {volume} {129}},\ \bibinfo {pages} {193604} (\bibinfo {year}
  {2022})}\BibitemShut {NoStop}%
\bibitem [{\citenamefont {Basso~Basset}\ \emph {et~al.}(2022)\citenamefont
  {Basso~Basset}, \citenamefont {Rota}, \citenamefont {Beccaceci},
  \citenamefont {Krieger}, \citenamefont {Buchinger}, \citenamefont {Neuwirth},
  \citenamefont {Huet}, \citenamefont {Stroj}, \citenamefont {da~Silva},
  \citenamefont {Schimpf} \emph {et~al.}}]{basso2023}%
  \BibitemOpen
  \bibfield  {author} {\bibinfo {author} {\bibfnamefont {F.}~\bibnamefont
  {Basso~Basset}}, \bibinfo {author} {\bibfnamefont {M.~B.}\ \bibnamefont
  {Rota}}, \bibinfo {author} {\bibfnamefont {M.}~\bibnamefont {Beccaceci}},
  \bibinfo {author} {\bibfnamefont {T.~M.}\ \bibnamefont {Krieger}}, \bibinfo
  {author} {\bibfnamefont {Q.}~\bibnamefont {Buchinger}}, \bibinfo {author}
  {\bibfnamefont {J.}~\bibnamefont {Neuwirth}}, \bibinfo {author}
  {\bibfnamefont {H.}~\bibnamefont {Huet}}, \bibinfo {author} {\bibfnamefont
  {S.}~\bibnamefont {Stroj}}, \bibinfo {author} {\bibfnamefont {S.~F.~C.}\
  \bibnamefont {da~Silva}}, \bibinfo {author} {\bibfnamefont {C.}~\bibnamefont
  {Schimpf}}, \emph {et~al.},\ }\bibfield  {title} {\bibinfo {title}
  {Signatures of the optical stark effect on entangled photon pairs from
  resonantly-pumped quantum dots},\ }\bibfield  {journal} {\bibinfo  {journal}
  {arXiv preprint arXiv:2212.07087}\ }\href
  {https://doi.org/10.48550/arXiv.2212.07087} {10.48550/arXiv.2212.07087}
  (\bibinfo {year} {2022})\BibitemShut {NoStop}%
\bibitem [{\citenamefont {Bracht}\ \emph {et~al.}(2021)\citenamefont {Bracht},
  \citenamefont {Cosacchi}, \citenamefont {Seidelmann}, \citenamefont
  {Cygorek}, \citenamefont {Vagov}, \citenamefont {Axt}, \citenamefont
  {Heindel},\ and\ \citenamefont {Reiter}}]{bracht21swingup}%
  \BibitemOpen
  \bibfield  {author} {\bibinfo {author} {\bibfnamefont {T.~K.}\ \bibnamefont
  {Bracht}}, \bibinfo {author} {\bibfnamefont {M.}~\bibnamefont {Cosacchi}},
  \bibinfo {author} {\bibfnamefont {T.}~\bibnamefont {Seidelmann}}, \bibinfo
  {author} {\bibfnamefont {M.}~\bibnamefont {Cygorek}}, \bibinfo {author}
  {\bibfnamefont {A.}~\bibnamefont {Vagov}}, \bibinfo {author} {\bibfnamefont
  {V.~M.}\ \bibnamefont {Axt}}, \bibinfo {author} {\bibfnamefont
  {T.}~\bibnamefont {Heindel}},\ and\ \bibinfo {author} {\bibfnamefont {D.~E.}\
  \bibnamefont {Reiter}},\ }\bibfield  {title} {\bibinfo {title} {Swing-up of
  quantum emitter population using detuned pulses},\ }\href
  {https://doi.org/10.1103/PRXQuantum.2.040354} {\bibfield  {journal} {\bibinfo
   {journal} {PRX Quantum}\ }\textbf {\bibinfo {volume} {2}},\ \bibinfo {pages}
  {040354} (\bibinfo {year} {2021})}\BibitemShut {NoStop}%
\bibitem [{\citenamefont {Karli}\ \emph {et~al.}(2022)\citenamefont {Karli},
  \citenamefont {Kappe}, \citenamefont {Remesh}, \citenamefont {Bracht},
  \citenamefont {Münzberg}, \citenamefont {Covre~da Silva}, \citenamefont
  {Seidelmann}, \citenamefont {Axt}, \citenamefont {Rastelli}, \citenamefont
  {Reiter},\ and\ \citenamefont {Weihs}}]{karli2022super}%
  \BibitemOpen
  \bibfield  {author} {\bibinfo {author} {\bibfnamefont {Y.}~\bibnamefont
  {Karli}}, \bibinfo {author} {\bibfnamefont {F.}~\bibnamefont {Kappe}},
  \bibinfo {author} {\bibfnamefont {V.}~\bibnamefont {Remesh}}, \bibinfo
  {author} {\bibfnamefont {T.~K.}\ \bibnamefont {Bracht}}, \bibinfo {author}
  {\bibfnamefont {J.}~\bibnamefont {Münzberg}}, \bibinfo {author}
  {\bibfnamefont {S.}~\bibnamefont {Covre~da Silva}}, \bibinfo {author}
  {\bibfnamefont {T.}~\bibnamefont {Seidelmann}}, \bibinfo {author}
  {\bibfnamefont {V.~M.}\ \bibnamefont {Axt}}, \bibinfo {author} {\bibfnamefont
  {A.}~\bibnamefont {Rastelli}}, \bibinfo {author} {\bibfnamefont {D.~E.}\
  \bibnamefont {Reiter}},\ and\ \bibinfo {author} {\bibfnamefont
  {G.}~\bibnamefont {Weihs}},\ }\bibfield  {title} {\bibinfo {title} {Super
  scheme in action: Experimental demonstration of red-detuned excitation of a
  quantum emitter},\ }\href {https://doi.org/10.1021/acs.nanolett.2c01783}
  {\bibfield  {journal} {\bibinfo  {journal} {Nano Lett.}\ }\textbf {\bibinfo
  {volume} {22}},\ \bibinfo {pages} {6567} (\bibinfo {year}
  {2022})}\BibitemShut {NoStop}%
\bibitem [{\citenamefont {Boos}\ \emph {et~al.}(2022)\citenamefont {Boos},
  \citenamefont {Sbresny}, \citenamefont {Kim}, \citenamefont {Kremser},
  \citenamefont {Riedl}, \citenamefont {Bopp}, \citenamefont {Rauhaus},
  \citenamefont {Scaparra}, \citenamefont {Jöns}, \citenamefont {Finley},
  \citenamefont {Müller},\ and\ \citenamefont {Hanschke}}]{boos2022coherent}%
  \BibitemOpen
  \bibfield  {author} {\bibinfo {author} {\bibfnamefont {K.}~\bibnamefont
  {Boos}}, \bibinfo {author} {\bibfnamefont {F.}~\bibnamefont {Sbresny}},
  \bibinfo {author} {\bibfnamefont {S.~K.}\ \bibnamefont {Kim}}, \bibinfo
  {author} {\bibfnamefont {M.}~\bibnamefont {Kremser}}, \bibinfo {author}
  {\bibfnamefont {H.}~\bibnamefont {Riedl}}, \bibinfo {author} {\bibfnamefont
  {F.~W.}\ \bibnamefont {Bopp}}, \bibinfo {author} {\bibfnamefont
  {W.}~\bibnamefont {Rauhaus}}, \bibinfo {author} {\bibfnamefont
  {B.}~\bibnamefont {Scaparra}}, \bibinfo {author} {\bibfnamefont {K.~D.}\
  \bibnamefont {Jöns}}, \bibinfo {author} {\bibfnamefont {J.~J.}\ \bibnamefont
  {Finley}}, \bibinfo {author} {\bibfnamefont {K.}~\bibnamefont {Müller}},\
  and\ \bibinfo {author} {\bibfnamefont {L.}~\bibnamefont {Hanschke}},\
  }\bibfield  {title} {\bibinfo {title} {Coherent dynamics of the swing-up
  excitation technique},\ }\href {https://arxiv.org/abs/2211.14289} {\bibfield
  {journal} {\bibinfo  {journal} {arXiv preprint arXiv:2211.14289}\ } (\bibinfo
  {year} {2022})}\BibitemShut {NoStop}%
\bibitem [{\citenamefont {Heinisch}\ \emph {et~al.}(2023)\citenamefont
  {Heinisch}, \citenamefont {K{\"o}cher}, \citenamefont {Bauch},\ and\
  \citenamefont {Schumacher}}]{heinisch2023arxiv}%
  \BibitemOpen
  \bibfield  {author} {\bibinfo {author} {\bibfnamefont {N.}~\bibnamefont
  {Heinisch}}, \bibinfo {author} {\bibfnamefont {N.}~\bibnamefont
  {K{\"o}cher}}, \bibinfo {author} {\bibfnamefont {D.}~\bibnamefont {Bauch}},\
  and\ \bibinfo {author} {\bibfnamefont {S.}~\bibnamefont {Schumacher}},\
  }\bibfield  {title} {\bibinfo {title} {Swing-up dynamics in quantum emitter
  cavity systems},\ }\bibfield  {journal} {\bibinfo  {journal} {arXiv preprint
  arXiv:2303.12604}\ }\href {https://doi.org/10.48550/arXiv.2303.12604}
  {10.48550/arXiv.2303.12604} (\bibinfo {year} {2023})\BibitemShut {NoStop}%
\bibitem [{\citenamefont {Ota}\ \emph {et~al.}(2011)\citenamefont {Ota},
  \citenamefont {Iwamoto}, \citenamefont {Kumagai},\ and\ \citenamefont
  {Arakawa}}]{ota2011spontaneous}%
  \BibitemOpen
  \bibfield  {author} {\bibinfo {author} {\bibfnamefont {Y.}~\bibnamefont
  {Ota}}, \bibinfo {author} {\bibfnamefont {S.}~\bibnamefont {Iwamoto}},
  \bibinfo {author} {\bibfnamefont {N.}~\bibnamefont {Kumagai}},\ and\ \bibinfo
  {author} {\bibfnamefont {Y.}~\bibnamefont {Arakawa}},\ }\bibfield  {title}
  {\bibinfo {title} {Spontaneous two-photon emission from a single quantum
  dot},\ }\href {https://doi.org/10.1103/PhysRevLett.107.233602} {\bibfield
  {journal} {\bibinfo  {journal} {Phys. Rev. Lett.}\ }\textbf {\bibinfo
  {volume} {107}},\ \bibinfo {pages} {233602} (\bibinfo {year}
  {2011})}\BibitemShut {NoStop}%
\bibitem [{\citenamefont {Reitzenstein}\ and\ \citenamefont
  {Forchel}(2010)}]{reitzenstein2010quantum}%
  \BibitemOpen
  \bibfield  {author} {\bibinfo {author} {\bibfnamefont {S.}~\bibnamefont
  {Reitzenstein}}\ and\ \bibinfo {author} {\bibfnamefont {A.}~\bibnamefont
  {Forchel}},\ }\bibfield  {title} {\bibinfo {title} {Quantum dot
  micropillars},\ }\href {https://doi.org/10.1088/0022-3727/43/3/033001}
  {\bibfield  {journal} {\bibinfo  {journal} {J. Phys. D: Appl. Phys.}\
  }\textbf {\bibinfo {volume} {43}},\ \bibinfo {pages} {033001} (\bibinfo
  {year} {2010})}\BibitemShut {NoStop}%
\bibitem [{\citenamefont {Lodahl}\ \emph {et~al.}(2015)\citenamefont {Lodahl},
  \citenamefont {Mahmoodian},\ and\ \citenamefont
  {Stobbe}}]{lodahl2015interfacing}%
  \BibitemOpen
  \bibfield  {author} {\bibinfo {author} {\bibfnamefont {P.}~\bibnamefont
  {Lodahl}}, \bibinfo {author} {\bibfnamefont {S.}~\bibnamefont {Mahmoodian}},\
  and\ \bibinfo {author} {\bibfnamefont {S.}~\bibnamefont {Stobbe}},\
  }\bibfield  {title} {\bibinfo {title} {Interfacing single photons and single
  quantum dots with photonic nanostructures},\ }\href
  {https://journals.aps.org/rmp/abstract/10.1103/RevModPhys.87.347} {\bibfield
  {journal} {\bibinfo  {journal} {Rev. Mod. Phys.}\ }\textbf {\bibinfo {volume}
  {87}},\ \bibinfo {pages} {347} (\bibinfo {year} {2015})}\BibitemShut
  {NoStop}%
\bibitem [{\citenamefont {Wang}\ \emph {et~al.}(2019)\citenamefont {Wang},
  \citenamefont {Hu}, \citenamefont {Chung}, \citenamefont {Qin}, \citenamefont
  {Yang}, \citenamefont {Li}, \citenamefont {Liu}, \citenamefont {Zhong},
  \citenamefont {He}, \citenamefont {Ding}, \citenamefont {Deng}, \citenamefont
  {Dai}, \citenamefont {Huo}, \citenamefont {H\"ofling}, \citenamefont {Lu},\
  and\ \citenamefont {Pan}}]{wang2019ondemand}%
  \BibitemOpen
  \bibfield  {author} {\bibinfo {author} {\bibfnamefont {H.}~\bibnamefont
  {Wang}}, \bibinfo {author} {\bibfnamefont {H.}~\bibnamefont {Hu}}, \bibinfo
  {author} {\bibfnamefont {T.-H.}\ \bibnamefont {Chung}}, \bibinfo {author}
  {\bibfnamefont {J.}~\bibnamefont {Qin}}, \bibinfo {author} {\bibfnamefont
  {X.}~\bibnamefont {Yang}}, \bibinfo {author} {\bibfnamefont {J.-P.}\
  \bibnamefont {Li}}, \bibinfo {author} {\bibfnamefont {R.-Z.}\ \bibnamefont
  {Liu}}, \bibinfo {author} {\bibfnamefont {H.-S.}\ \bibnamefont {Zhong}},
  \bibinfo {author} {\bibfnamefont {Y.-M.}\ \bibnamefont {He}}, \bibinfo
  {author} {\bibfnamefont {X.}~\bibnamefont {Ding}}, \bibinfo {author}
  {\bibfnamefont {Y.-H.}\ \bibnamefont {Deng}}, \bibinfo {author}
  {\bibfnamefont {Q.}~\bibnamefont {Dai}}, \bibinfo {author} {\bibfnamefont
  {Y.-H.}\ \bibnamefont {Huo}}, \bibinfo {author} {\bibfnamefont
  {S.}~\bibnamefont {H\"ofling}}, \bibinfo {author} {\bibfnamefont {C.-Y.}\
  \bibnamefont {Lu}},\ and\ \bibinfo {author} {\bibfnamefont {J.-W.}\
  \bibnamefont {Pan}},\ }\bibfield  {title} {\bibinfo {title} {On-demand
  semiconductor source of entangled photons which simultaneously has high
  fidelity, efficiency, and indistinguishability},\ }\href
  {https://doi.org/10.1103/PhysRevLett.122.113602} {\bibfield  {journal}
  {\bibinfo  {journal} {Phys. Rev. Lett.}\ }\textbf {\bibinfo {volume} {122}},\
  \bibinfo {pages} {113602} (\bibinfo {year} {2019})}\BibitemShut {NoStop}%
\bibitem [{\citenamefont {Rickert}\ \emph {et~al.}(2023)\citenamefont
  {Rickert}, \citenamefont {Betz}, \citenamefont {Plock}, \citenamefont
  {Burger},\ and\ \citenamefont {Heindel}}]{rickert2023high}%
  \BibitemOpen
  \bibfield  {author} {\bibinfo {author} {\bibfnamefont {L.}~\bibnamefont
  {Rickert}}, \bibinfo {author} {\bibfnamefont {F.}~\bibnamefont {Betz}},
  \bibinfo {author} {\bibfnamefont {M.}~\bibnamefont {Plock}}, \bibinfo
  {author} {\bibfnamefont {S.}~\bibnamefont {Burger}},\ and\ \bibinfo {author}
  {\bibfnamefont {T.}~\bibnamefont {Heindel}},\ }\bibfield  {title} {\bibinfo
  {title} {High-performance designs for fiber-pigtailed quantum-light sources
  based on quantum dots in electrically-controlled circular bragg gratings},\
  }\href {https://opg.optica.org/oe/fulltext.cfm?uri=oe-31-9-14750&id=529227}
  {\bibfield  {journal} {\bibinfo  {journal} {Opt. Express}\ }\textbf {\bibinfo
  {volume} {31}},\ \bibinfo {pages} {14750} (\bibinfo {year}
  {2023})}\BibitemShut {NoStop}%
\bibitem [{\citenamefont {Ramsay}\ \emph {et~al.}(2010)\citenamefont {Ramsay},
  \citenamefont {Gopal}, \citenamefont {Gauger}, \citenamefont {Nazir},
  \citenamefont {Lovett}, \citenamefont {Fox},\ and\ \citenamefont
  {Skolnick}}]{ramsay2010phonon}%
  \BibitemOpen
  \bibfield  {author} {\bibinfo {author} {\bibfnamefont {A.~J.}\ \bibnamefont
  {Ramsay}}, \bibinfo {author} {\bibfnamefont {A.~V.}\ \bibnamefont {Gopal}},
  \bibinfo {author} {\bibfnamefont {E.~M.}\ \bibnamefont {Gauger}}, \bibinfo
  {author} {\bibfnamefont {A.}~\bibnamefont {Nazir}}, \bibinfo {author}
  {\bibfnamefont {B.~W.}\ \bibnamefont {Lovett}}, \bibinfo {author}
  {\bibfnamefont {A.~M.}\ \bibnamefont {Fox}},\ and\ \bibinfo {author}
  {\bibfnamefont {M.~S.}\ \bibnamefont {Skolnick}},\ }\bibfield  {title}
  {\bibinfo {title} {Damping of exciton rabi rotations by acoustic phonons in
  optically excited $\mathrm{InGaAs}/\mathrm{GaAs}$ quantum dots},\ }\href
  {https://doi.org/10.1103/PhysRevLett.104.017402} {\bibfield  {journal}
  {\bibinfo  {journal} {Phys. Rev. Lett.}\ }\textbf {\bibinfo {volume} {104}},\
  \bibinfo {pages} {017402} (\bibinfo {year} {2010})}\BibitemShut {NoStop}%
\bibitem [{\citenamefont {L{\"u}ker}\ and\ \citenamefont
  {Reiter}(2019)}]{luker2019review}%
  \BibitemOpen
  \bibfield  {author} {\bibinfo {author} {\bibfnamefont {S.}~\bibnamefont
  {L{\"u}ker}}\ and\ \bibinfo {author} {\bibfnamefont {D.~E.}\ \bibnamefont
  {Reiter}},\ }\bibfield  {title} {\bibinfo {title} {A review on optical
  excitation of semiconductor quantum dots under the influence of phonons},\
  }\href {https://doi.org/10.1088/1361-6641/ab1c14} {\bibfield  {journal}
  {\bibinfo  {journal} {Semicond. Sci. Technol.}\ }\textbf {\bibinfo {volume}
  {34}},\ \bibinfo {pages} {063002} (\bibinfo {year} {2019})}\BibitemShut
  {NoStop}%
\bibitem [{\citenamefont {Cosacchi}\ \emph {et~al.}(2021)\citenamefont
  {Cosacchi}, \citenamefont {Seidelmann}, \citenamefont {Cygorek},
  \citenamefont {Vagov}, \citenamefont {Reiter},\ and\ \citenamefont
  {Axt}}]{cosacchi2021accuracy}%
  \BibitemOpen
  \bibfield  {author} {\bibinfo {author} {\bibfnamefont {M.}~\bibnamefont
  {Cosacchi}}, \bibinfo {author} {\bibfnamefont {T.}~\bibnamefont
  {Seidelmann}}, \bibinfo {author} {\bibfnamefont {M.}~\bibnamefont {Cygorek}},
  \bibinfo {author} {\bibfnamefont {A.}~\bibnamefont {Vagov}}, \bibinfo
  {author} {\bibfnamefont {D.~E.}\ \bibnamefont {Reiter}},\ and\ \bibinfo
  {author} {\bibfnamefont {V.~M.}\ \bibnamefont {Axt}},\ }\bibfield  {title}
  {\bibinfo {title} {Accuracy of the quantum regression theorem for photon
  emission from a quantum dot},\ }\href
  {https://doi.org/10.1103/PhysRevLett.127.100402} {\bibfield  {journal}
  {\bibinfo  {journal} {Phys. Rev. Lett.}\ }\textbf {\bibinfo {volume} {127}},\
  \bibinfo {pages} {100402} (\bibinfo {year} {2021})}\BibitemShut {NoStop}%
\bibitem [{\citenamefont {Vannucci}\ and\ \citenamefont
  {Gregersen}(2023)}]{vannucci2022phonon}%
  \BibitemOpen
  \bibfield  {author} {\bibinfo {author} {\bibfnamefont {L.}~\bibnamefont
  {Vannucci}}\ and\ \bibinfo {author} {\bibfnamefont {N.}~\bibnamefont
  {Gregersen}},\ }\bibfield  {title} {\bibinfo {title} {Highly efficient and
  indistinguishable single-photon sources via phonon-decoupled two-color
  excitation},\ }\href {https://doi.org/10.1103/PhysRevB.107.195306} {\bibfield
   {journal} {\bibinfo  {journal} {Phys. Rev. B}\ }\textbf {\bibinfo {volume}
  {107}},\ \bibinfo {pages} {195306} (\bibinfo {year} {2023})}\BibitemShut
  {NoStop}%
\bibitem [{\citenamefont {Carmele}\ and\ \citenamefont
  {Knorr}(2011)}]{carmele}%
  \BibitemOpen
  \bibfield  {author} {\bibinfo {author} {\bibfnamefont {A.}~\bibnamefont
  {Carmele}}\ and\ \bibinfo {author} {\bibfnamefont {A.}~\bibnamefont
  {Knorr}},\ }\bibfield  {title} {\bibinfo {title} {Analytical solution of the
  quantum-state tomography of the biexciton cascade in semiconductor quantum
  dots: Pure dephasing does not affect entanglement},\ }\href
  {https://doi.org/10.1103/PhysRevB.84.075328} {\bibfield  {journal} {\bibinfo
  {journal} {Phys. Rev. B}\ }\textbf {\bibinfo {volume} {84}},\ \bibinfo
  {pages} {075328} (\bibinfo {year} {2011})}\BibitemShut {NoStop}%
\bibitem [{\citenamefont {Seidelmann}\ \emph
  {et~al.}(2019{\natexlab{a}})\citenamefont {Seidelmann}, \citenamefont
  {Ungar}, \citenamefont {Cygorek}, \citenamefont {Vagov}, \citenamefont
  {Barth}, \citenamefont {Kuhn},\ and\ \citenamefont
  {Axt}}]{seidelmann2019from}%
  \BibitemOpen
  \bibfield  {author} {\bibinfo {author} {\bibfnamefont {T.}~\bibnamefont
  {Seidelmann}}, \bibinfo {author} {\bibfnamefont {F.}~\bibnamefont {Ungar}},
  \bibinfo {author} {\bibfnamefont {M.}~\bibnamefont {Cygorek}}, \bibinfo
  {author} {\bibfnamefont {A.}~\bibnamefont {Vagov}}, \bibinfo {author}
  {\bibfnamefont {A.~M.}\ \bibnamefont {Barth}}, \bibinfo {author}
  {\bibfnamefont {T.}~\bibnamefont {Kuhn}},\ and\ \bibinfo {author}
  {\bibfnamefont {V.~M.}\ \bibnamefont {Axt}},\ }\bibfield  {title} {\bibinfo
  {title} {From strong to weak temperature dependence of the two-photon
  entanglement resulting from the biexciton cascade inside a cavity},\ }\href
  {https://doi.org/10.1103/PhysRevB.99.245301} {\bibfield  {journal} {\bibinfo
  {journal} {Phys. Rev. B}\ }\textbf {\bibinfo {volume} {99}},\ \bibinfo
  {pages} {245301} (\bibinfo {year} {2019}{\natexlab{a}})}\BibitemShut
  {NoStop}%
\bibitem [{\citenamefont {Heinze}\ \emph {et~al.}(2017)\citenamefont {Heinze},
  \citenamefont {Zrenner},\ and\ \citenamefont
  {Schumacher}}]{heinze2017polarization}%
  \BibitemOpen
  \bibfield  {author} {\bibinfo {author} {\bibfnamefont {D.}~\bibnamefont
  {Heinze}}, \bibinfo {author} {\bibfnamefont {A.}~\bibnamefont {Zrenner}},\
  and\ \bibinfo {author} {\bibfnamefont {S.}~\bibnamefont {Schumacher}},\
  }\bibfield  {title} {\bibinfo {title} {Polarization-entangled twin photons
  from two-photon quantum-dot emission},\ }\href
  {https://doi.org/10.1103/PhysRevB.95.245306} {\bibfield  {journal} {\bibinfo
  {journal} {Phys. Rev. B}\ }\textbf {\bibinfo {volume} {95}},\ \bibinfo
  {pages} {245306} (\bibinfo {year} {2017})}\BibitemShut {NoStop}%
\bibitem [{\citenamefont {Seidelmann}\ \emph {et~al.}(2023)\citenamefont
  {Seidelmann}, \citenamefont {Bracht}, \citenamefont {Lehner}, \citenamefont
  {Schimpf}, \citenamefont {Cosacchi}, \citenamefont {Cygorek}, \citenamefont
  {Vagov}, \citenamefont {Rastelli}, \citenamefont {Reiter},\ and\
  \citenamefont {Axt}}]{Seidelmann2023}%
  \BibitemOpen
  \bibfield  {author} {\bibinfo {author} {\bibfnamefont {T.}~\bibnamefont
  {Seidelmann}}, \bibinfo {author} {\bibfnamefont {T.~K.}\ \bibnamefont
  {Bracht}}, \bibinfo {author} {\bibfnamefont {B.~U.}\ \bibnamefont {Lehner}},
  \bibinfo {author} {\bibfnamefont {C.}~\bibnamefont {Schimpf}}, \bibinfo
  {author} {\bibfnamefont {M.}~\bibnamefont {Cosacchi}}, \bibinfo {author}
  {\bibfnamefont {M.}~\bibnamefont {Cygorek}}, \bibinfo {author} {\bibfnamefont
  {A.}~\bibnamefont {Vagov}}, \bibinfo {author} {\bibfnamefont
  {A.}~\bibnamefont {Rastelli}}, \bibinfo {author} {\bibfnamefont {D.~E.}\
  \bibnamefont {Reiter}},\ and\ \bibinfo {author} {\bibfnamefont {V.~M.}\
  \bibnamefont {Axt}},\ }\bibfield  {title} {\bibinfo {title} {Two-photon
  excitation with finite pulses unlocks pure dephasing-induced degradation of
  entangled photons emitted by quantum dots},\ }\href
  {https://doi.org/10.1103/PhysRevB.107.235304} {\bibfield  {journal} {\bibinfo
   {journal} {Phys. Rev. B}\ }\textbf {\bibinfo {volume} {107}},\ \bibinfo
  {pages} {235304} (\bibinfo {year} {2023})}\BibitemShut {NoStop}%
\bibitem [{\citenamefont {Lehner}\ \emph {et~al.}(2023)\citenamefont {Lehner},
  \citenamefont {Seidelmann}, \citenamefont {Undeutsch}, \citenamefont
  {Schimpf}, \citenamefont {Manna}, \citenamefont {Gawe{\l}czyk}, \citenamefont
  {Covre~da Silva}, \citenamefont {Yuan}, \citenamefont {Stroj}, \citenamefont
  {Reiter}, \citenamefont {Axt},\ and\ \citenamefont
  {Rastelli}}]{lehner2023beyond}%
  \BibitemOpen
  \bibfield  {author} {\bibinfo {author} {\bibfnamefont {B.~U.}\ \bibnamefont
  {Lehner}}, \bibinfo {author} {\bibfnamefont {T.}~\bibnamefont {Seidelmann}},
  \bibinfo {author} {\bibfnamefont {G.}~\bibnamefont {Undeutsch}}, \bibinfo
  {author} {\bibfnamefont {C.}~\bibnamefont {Schimpf}}, \bibinfo {author}
  {\bibfnamefont {S.}~\bibnamefont {Manna}}, \bibinfo {author} {\bibfnamefont
  {M.}~\bibnamefont {Gawe{\l}czyk}}, \bibinfo {author} {\bibfnamefont {S.~F.}\
  \bibnamefont {Covre~da Silva}}, \bibinfo {author} {\bibfnamefont
  {X.}~\bibnamefont {Yuan}}, \bibinfo {author} {\bibfnamefont {S.}~\bibnamefont
  {Stroj}}, \bibinfo {author} {\bibfnamefont {D.~E.}\ \bibnamefont {Reiter}},
  \bibinfo {author} {\bibfnamefont {V.~M.}\ \bibnamefont {Axt}},\ and\ \bibinfo
  {author} {\bibfnamefont {A.}~\bibnamefont {Rastelli}},\ }\bibfield  {title}
  {\bibinfo {title} {Beyond the four-level model: Dark and hot states in
  quantum dots degrade photonic entanglement},\ }\href
  {https://doi.org/10.1021/acs.nanolett.2c04734} {\bibfield  {journal}
  {\bibinfo  {journal} {Nano Letters}\ }\textbf {\bibinfo {volume} {23}},\
  \bibinfo {pages} {1409} (\bibinfo {year} {2023})}\BibitemShut {NoStop}%
\bibitem [{\citenamefont {Yin}\ \emph {et~al.}(2020)\citenamefont {Yin},
  \citenamefont {Li}, \citenamefont {Liao}, \citenamefont {Yang}, \citenamefont
  {Cao}, \citenamefont {Zhang}, \citenamefont {Ren}, \citenamefont {Cai},
  \citenamefont {Liu}, \citenamefont {Li}, \citenamefont {Shu}, \citenamefont
  {Huang}, \citenamefont {Deng}, \citenamefont {Li}, \citenamefont {Zhang},
  \citenamefont {Liu}, \citenamefont {Chen}, \citenamefont {Lu}, \citenamefont
  {Wang}, \citenamefont {Xu}, \citenamefont {Wang}, \citenamefont {Peng},
  \citenamefont {Ekert},\ and\ \citenamefont {Pan}}]{yin2020entanglement}%
  \BibitemOpen
  \bibfield  {author} {\bibinfo {author} {\bibfnamefont {J.}~\bibnamefont
  {Yin}}, \bibinfo {author} {\bibfnamefont {Y.-H.}\ \bibnamefont {Li}},
  \bibinfo {author} {\bibfnamefont {S.-K.}\ \bibnamefont {Liao}}, \bibinfo
  {author} {\bibfnamefont {M.}~\bibnamefont {Yang}}, \bibinfo {author}
  {\bibfnamefont {Y.}~\bibnamefont {Cao}}, \bibinfo {author} {\bibfnamefont
  {L.}~\bibnamefont {Zhang}}, \bibinfo {author} {\bibfnamefont {J.-G.}\
  \bibnamefont {Ren}}, \bibinfo {author} {\bibfnamefont {W.-Q.}\ \bibnamefont
  {Cai}}, \bibinfo {author} {\bibfnamefont {W.-Y.}\ \bibnamefont {Liu}},
  \bibinfo {author} {\bibfnamefont {S.-L.}\ \bibnamefont {Li}}, \bibinfo
  {author} {\bibfnamefont {R.}~\bibnamefont {Shu}}, \bibinfo {author}
  {\bibfnamefont {Y.-M.}\ \bibnamefont {Huang}}, \bibinfo {author}
  {\bibfnamefont {L.}~\bibnamefont {Deng}}, \bibinfo {author} {\bibfnamefont
  {L.}~\bibnamefont {Li}}, \bibinfo {author} {\bibfnamefont {Q.}~\bibnamefont
  {Zhang}}, \bibinfo {author} {\bibfnamefont {N.-L.}\ \bibnamefont {Liu}},
  \bibinfo {author} {\bibfnamefont {Y.-A.}\ \bibnamefont {Chen}}, \bibinfo
  {author} {\bibfnamefont {C.-Y.}\ \bibnamefont {Lu}}, \bibinfo {author}
  {\bibfnamefont {X.-B.}\ \bibnamefont {Wang}}, \bibinfo {author}
  {\bibfnamefont {F.}~\bibnamefont {Xu}}, \bibinfo {author} {\bibfnamefont
  {J.-Y.}\ \bibnamefont {Wang}}, \bibinfo {author} {\bibfnamefont {C.-Z.}\
  \bibnamefont {Peng}}, \bibinfo {author} {\bibfnamefont {A.~K.}\ \bibnamefont
  {Ekert}},\ and\ \bibinfo {author} {\bibfnamefont {J.-W.}\ \bibnamefont
  {Pan}},\ }\bibfield  {title} {\bibinfo {title} {Entanglement-based secure
  quantum cryptography over 1,120 kilometres},\ }\href
  {https://doi.org/10.1038/s41586-020-2401-y} {\bibfield  {journal} {\bibinfo
  {journal} {Nature}\ }\textbf {\bibinfo {volume} {582}},\ \bibinfo {pages}
  {501} (\bibinfo {year} {2020})}\BibitemShut {NoStop}%
\bibitem [{\citenamefont {Reiter}\ \emph {et~al.}(2019)\citenamefont {Reiter},
  \citenamefont {Kuhn},\ and\ \citenamefont {Axt}}]{reiter2019distinctive}%
  \BibitemOpen
  \bibfield  {author} {\bibinfo {author} {\bibfnamefont {D.~E.}\ \bibnamefont
  {Reiter}}, \bibinfo {author} {\bibfnamefont {T.}~\bibnamefont {Kuhn}},\ and\
  \bibinfo {author} {\bibfnamefont {V.~M.}\ \bibnamefont {Axt}},\ }\bibfield
  {title} {\bibinfo {title} {Distinctive characteristics of carrier-phonon
  interactions in optically driven semiconductor quantum dots},\ }\href
  {https://doi.org/10.1080/23746149.2019.1655478} {\bibfield  {journal}
  {\bibinfo  {journal} {Adv. Phys.: X}\ }\textbf {\bibinfo {volume} {4}},\
  \bibinfo {pages} {1655478} (\bibinfo {year} {2019})}\BibitemShut {NoStop}%
\bibitem [{\citenamefont {Cygorek}\ \emph {et~al.}(2022)\citenamefont
  {Cygorek}, \citenamefont {Cosacchi}, \citenamefont {Vagov}, \citenamefont
  {Axt}, \citenamefont {Lovett}, \citenamefont {Keeling},\ and\ \citenamefont
  {Gauger}}]{cygorek2022simulation}%
  \BibitemOpen
  \bibfield  {author} {\bibinfo {author} {\bibfnamefont {M.}~\bibnamefont
  {Cygorek}}, \bibinfo {author} {\bibfnamefont {M.}~\bibnamefont {Cosacchi}},
  \bibinfo {author} {\bibfnamefont {A.}~\bibnamefont {Vagov}}, \bibinfo
  {author} {\bibfnamefont {V.~M.}\ \bibnamefont {Axt}}, \bibinfo {author}
  {\bibfnamefont {B.~W.}\ \bibnamefont {Lovett}}, \bibinfo {author}
  {\bibfnamefont {J.}~\bibnamefont {Keeling}},\ and\ \bibinfo {author}
  {\bibfnamefont {E.~M.}\ \bibnamefont {Gauger}},\ }\bibfield  {title}
  {\bibinfo {title} {Simulation of open quantum systems by automated
  compression of arbitrary environments},\ }\href
  {https://doi.org/10.1038/s41567-022-01544-9} {\bibfield  {journal} {\bibinfo
  {journal} {Nature Physics}\ }\textbf {\bibinfo {volume} {18}},\ \bibinfo
  {pages} {662} (\bibinfo {year} {2022})}\BibitemShut {NoStop}%
\bibitem [{\citenamefont {Strathearn}\ \emph {et~al.}(2018)\citenamefont
  {Strathearn}, \citenamefont {Kirton}, \citenamefont {Kilda}, \citenamefont
  {Keeling},\ and\ \citenamefont {Lovett}}]{strathearn2018efficient}%
  \BibitemOpen
  \bibfield  {author} {\bibinfo {author} {\bibfnamefont {A.}~\bibnamefont
  {Strathearn}}, \bibinfo {author} {\bibfnamefont {P.}~\bibnamefont {Kirton}},
  \bibinfo {author} {\bibfnamefont {D.}~\bibnamefont {Kilda}}, \bibinfo
  {author} {\bibfnamefont {J.}~\bibnamefont {Keeling}},\ and\ \bibinfo {author}
  {\bibfnamefont {B.~W.}\ \bibnamefont {Lovett}},\ }\bibfield  {title}
  {\bibinfo {title} {Efficient non-markovian quantum dynamics using
  time-evolving matrix product operators},\ }\href
  {https://doi.org/10.1038/s41467-018-05617-3} {\bibfield  {journal} {\bibinfo
  {journal} {Nature communications}\ }\textbf {\bibinfo {volume} {9}},\
  \bibinfo {pages} {3322} (\bibinfo {year} {2018})}\BibitemShut {NoStop}%
\bibitem [{\citenamefont {J\o{}rgensen}\ and\ \citenamefont
  {Pollock}(2019)}]{jorgensen2019exploiting}%
  \BibitemOpen
  \bibfield  {author} {\bibinfo {author} {\bibfnamefont {M.~R.}\ \bibnamefont
  {J\o{}rgensen}}\ and\ \bibinfo {author} {\bibfnamefont {F.~A.}\ \bibnamefont
  {Pollock}},\ }\bibfield  {title} {\bibinfo {title} {Exploiting the causal
  tensor network structure of quantum processes to efficiently simulate
  non-markovian path integrals},\ }\href
  {https://doi.org/10.1103/PhysRevLett.123.240602} {\bibfield  {journal}
  {\bibinfo  {journal} {Phys. Rev. Lett.}\ }\textbf {\bibinfo {volume} {123}},\
  \bibinfo {pages} {240602} (\bibinfo {year} {2019})}\BibitemShut {NoStop}%
\bibitem [{\citenamefont {Pollock}\ \emph {et~al.}(2018)\citenamefont
  {Pollock}, \citenamefont {Rodr\'{\i}guez-Rosario}, \citenamefont
  {Frauenheim}, \citenamefont {Paternostro},\ and\ \citenamefont
  {Modi}}]{pollock2018non}%
  \BibitemOpen
  \bibfield  {author} {\bibinfo {author} {\bibfnamefont {F.~A.}\ \bibnamefont
  {Pollock}}, \bibinfo {author} {\bibfnamefont {C.}~\bibnamefont
  {Rodr\'{\i}guez-Rosario}}, \bibinfo {author} {\bibfnamefont {T.}~\bibnamefont
  {Frauenheim}}, \bibinfo {author} {\bibfnamefont {M.}~\bibnamefont
  {Paternostro}},\ and\ \bibinfo {author} {\bibfnamefont {K.}~\bibnamefont
  {Modi}},\ }\bibfield  {title} {\bibinfo {title} {Non-markovian quantum
  processes: Complete framework and efficient characterization},\ }\href
  {https://doi.org/10.1103/PhysRevA.97.012127} {\bibfield  {journal} {\bibinfo
  {journal} {Phys. Rev. A}\ }\textbf {\bibinfo {volume} {97}},\ \bibinfo
  {pages} {012127} (\bibinfo {year} {2018})}\BibitemShut {NoStop}%
\bibitem [{\citenamefont {Cygorek}\ \emph {et~al.}(2023)\citenamefont
  {Cygorek}, \citenamefont {Keeling}, \citenamefont {Lovett},\ and\
  \citenamefont {Gauger}}]{cygorek2023sublinear}%
  \BibitemOpen
  \bibfield  {author} {\bibinfo {author} {\bibfnamefont {M.}~\bibnamefont
  {Cygorek}}, \bibinfo {author} {\bibfnamefont {J.}~\bibnamefont {Keeling}},
  \bibinfo {author} {\bibfnamefont {B.~W.}\ \bibnamefont {Lovett}},\ and\
  \bibinfo {author} {\bibfnamefont {E.~M.}\ \bibnamefont {Gauger}},\ }\bibfield
   {title} {\bibinfo {title} {Sublinear scaling in non-markovian open quantum
  systems simulations},\ }\bibfield  {journal} {\bibinfo  {journal} {arXiv
  preprint 2304.05291}\ }\href {https://doi.org/10.48550/arXiv.2304.05291}
  {10.48550/arXiv.2304.05291} (\bibinfo {year} {2023})\BibitemShut {NoStop}%
\bibitem [{\citenamefont {Cosacchi}\ \emph {et~al.}(2018)\citenamefont
  {Cosacchi}, \citenamefont {Cygorek}, \citenamefont {Ungar}, \citenamefont
  {Barth}, \citenamefont {Vagov},\ and\ \citenamefont
  {Axt}}]{cosacchi2018pathintegral}%
  \BibitemOpen
  \bibfield  {author} {\bibinfo {author} {\bibfnamefont {M.}~\bibnamefont
  {Cosacchi}}, \bibinfo {author} {\bibfnamefont {M.}~\bibnamefont {Cygorek}},
  \bibinfo {author} {\bibfnamefont {F.}~\bibnamefont {Ungar}}, \bibinfo
  {author} {\bibfnamefont {A.~M.}\ \bibnamefont {Barth}}, \bibinfo {author}
  {\bibfnamefont {A.}~\bibnamefont {Vagov}},\ and\ \bibinfo {author}
  {\bibfnamefont {V.~M.}\ \bibnamefont {Axt}},\ }\bibfield  {title} {\bibinfo
  {title} {Path-integral approach for nonequilibrium multitime correlation
  functions of open quantum systems coupled to markovian and non-markovian
  environments},\ }\href {https://doi.org/10.1103/PhysRevB.98.125302}
  {\bibfield  {journal} {\bibinfo  {journal} {Phys. Rev. B}\ }\textbf {\bibinfo
  {volume} {98}},\ \bibinfo {pages} {125302} (\bibinfo {year}
  {2018})}\BibitemShut {NoStop}%
\bibitem [{\citenamefont {Johansson}\ \emph {et~al.}(2012)\citenamefont
  {Johansson}, \citenamefont {Nation},\ and\ \citenamefont
  {Nori}}]{johansson2012qutip}%
  \BibitemOpen
  \bibfield  {author} {\bibinfo {author} {\bibfnamefont {J.}~\bibnamefont
  {Johansson}}, \bibinfo {author} {\bibfnamefont {P.}~\bibnamefont {Nation}},\
  and\ \bibinfo {author} {\bibfnamefont {F.}~\bibnamefont {Nori}},\ }\bibfield
  {title} {\bibinfo {title} {Qutip: An open-source python framework for the
  dynamics of open quantum systems},\ }\href
  {https://doi.org/https://doi.org/10.1016/j.cpc.2012.02.021} {\bibfield
  {journal} {\bibinfo  {journal} {Computer Physics Communications}\ }\textbf
  {\bibinfo {volume} {183}},\ \bibinfo {pages} {1760} (\bibinfo {year}
  {2012})}\BibitemShut {NoStop}%
\bibitem [{\citenamefont {Johansson}\ \emph {et~al.}(2013)\citenamefont
  {Johansson}, \citenamefont {Nation},\ and\ \citenamefont
  {Nori}}]{johansson2013qutip}%
  \BibitemOpen
  \bibfield  {author} {\bibinfo {author} {\bibfnamefont {J.}~\bibnamefont
  {Johansson}}, \bibinfo {author} {\bibfnamefont {P.}~\bibnamefont {Nation}},\
  and\ \bibinfo {author} {\bibfnamefont {F.}~\bibnamefont {Nori}},\ }\bibfield
  {title} {\bibinfo {title} {Qutip 2: A python framework for the dynamics of
  open quantum systems},\ }\href
  {https://doi.org/https://doi.org/10.1016/j.cpc.2012.11.019} {\bibfield
  {journal} {\bibinfo  {journal} {Computer Physics Communications}\ }\textbf
  {\bibinfo {volume} {184}},\ \bibinfo {pages} {1234} (\bibinfo {year}
  {2013})}\BibitemShut {NoStop}%
\bibitem [{\citenamefont {Schumacher}\ \emph {et~al.}(2012)\citenamefont
  {Schumacher}, \citenamefont {F\"{o}rstner}, \citenamefont {Zrenner},
  \citenamefont {Florian}, \citenamefont {Gies}, \citenamefont {Gartner},\ and\
  \citenamefont {Jahnke}}]{schumacher2012cavity}%
  \BibitemOpen
  \bibfield  {author} {\bibinfo {author} {\bibfnamefont {S.}~\bibnamefont
  {Schumacher}}, \bibinfo {author} {\bibfnamefont {J.}~\bibnamefont
  {F\"{o}rstner}}, \bibinfo {author} {\bibfnamefont {A.}~\bibnamefont
  {Zrenner}}, \bibinfo {author} {\bibfnamefont {M.}~\bibnamefont {Florian}},
  \bibinfo {author} {\bibfnamefont {C.}~\bibnamefont {Gies}}, \bibinfo {author}
  {\bibfnamefont {P.}~\bibnamefont {Gartner}},\ and\ \bibinfo {author}
  {\bibfnamefont {F.}~\bibnamefont {Jahnke}},\ }\bibfield  {title} {\bibinfo
  {title} {Cavity-assisted emission of polarization-entangled photons from
  biexcitons in quantum dots with fine-structure splitting},\ }\href
  {https://doi.org/10.1364/OE.20.005335} {\bibfield  {journal} {\bibinfo
  {journal} {Opt. Express}\ }\textbf {\bibinfo {volume} {20}},\ \bibinfo
  {pages} {5335} (\bibinfo {year} {2012})}\BibitemShut {NoStop}%
\bibitem [{\citenamefont {Seidelmann}\ \emph
  {et~al.}(2019{\natexlab{b}})\citenamefont {Seidelmann}, \citenamefont
  {Ungar}, \citenamefont {Barth}, \citenamefont {Vagov}, \citenamefont {Axt},
  \citenamefont {Cygorek},\ and\ \citenamefont {Kuhn}}]{seidelmann2019phonon}%
  \BibitemOpen
  \bibfield  {author} {\bibinfo {author} {\bibfnamefont {T.}~\bibnamefont
  {Seidelmann}}, \bibinfo {author} {\bibfnamefont {F.}~\bibnamefont {Ungar}},
  \bibinfo {author} {\bibfnamefont {A.~M.}\ \bibnamefont {Barth}}, \bibinfo
  {author} {\bibfnamefont {A.}~\bibnamefont {Vagov}}, \bibinfo {author}
  {\bibfnamefont {V.~M.}\ \bibnamefont {Axt}}, \bibinfo {author} {\bibfnamefont
  {M.}~\bibnamefont {Cygorek}},\ and\ \bibinfo {author} {\bibfnamefont
  {T.}~\bibnamefont {Kuhn}},\ }\bibfield  {title} {\bibinfo {title}
  {Phonon-induced enhancement of photon entanglement in quantum dot-cavity
  systems},\ }\href {https://doi.org/10.1103/PhysRevLett.123.137401} {\bibfield
   {journal} {\bibinfo  {journal} {Phys. Rev. Lett.}\ }\textbf {\bibinfo
  {volume} {123}},\ \bibinfo {pages} {137401} (\bibinfo {year}
  {2019}{\natexlab{b}})}\BibitemShut {NoStop}%
\bibitem [{\citenamefont {Vagov}\ \emph {et~al.}(2007)\citenamefont {Vagov},
  \citenamefont {Croitoru}, \citenamefont {Axt}, \citenamefont {Kuhn},\ and\
  \citenamefont {Peeters}}]{vagov2007nonmonotonic}%
  \BibitemOpen
  \bibfield  {author} {\bibinfo {author} {\bibfnamefont {A.}~\bibnamefont
  {Vagov}}, \bibinfo {author} {\bibfnamefont {M.~D.}\ \bibnamefont {Croitoru}},
  \bibinfo {author} {\bibfnamefont {V.~M.}\ \bibnamefont {Axt}}, \bibinfo
  {author} {\bibfnamefont {T.}~\bibnamefont {Kuhn}},\ and\ \bibinfo {author}
  {\bibfnamefont {F.~M.}\ \bibnamefont {Peeters}},\ }\bibfield  {title}
  {\bibinfo {title} {Nonmonotonic field dependence of damping and reappearance
  of rabi oscillations in quantum dots},\ }\href
  {https://doi.org/10.1103/PhysRevLett.98.227403} {\bibfield  {journal}
  {\bibinfo  {journal} {Phys. Rev. Lett.}\ }\textbf {\bibinfo {volume} {98}},\
  \bibinfo {pages} {227403} (\bibinfo {year} {2007})}\BibitemShut {NoStop}%
\bibitem [{\citenamefont {Kaldewey}\ \emph {et~al.}(2017)\citenamefont
  {Kaldewey}, \citenamefont {L\"uker}, \citenamefont {Kuhlmann}, \citenamefont
  {Valentin}, \citenamefont {Chauveau}, \citenamefont {Ludwig}, \citenamefont
  {Wieck}, \citenamefont {Reiter}, \citenamefont {Kuhn},\ and\ \citenamefont
  {Warburton}}]{kaldewey2017demonstrating}%
  \BibitemOpen
  \bibfield  {author} {\bibinfo {author} {\bibfnamefont {T.}~\bibnamefont
  {Kaldewey}}, \bibinfo {author} {\bibfnamefont {S.}~\bibnamefont {L\"uker}},
  \bibinfo {author} {\bibfnamefont {A.~V.}\ \bibnamefont {Kuhlmann}}, \bibinfo
  {author} {\bibfnamefont {S.~R.}\ \bibnamefont {Valentin}}, \bibinfo {author}
  {\bibfnamefont {J.-M.}\ \bibnamefont {Chauveau}}, \bibinfo {author}
  {\bibfnamefont {A.}~\bibnamefont {Ludwig}}, \bibinfo {author} {\bibfnamefont
  {A.~D.}\ \bibnamefont {Wieck}}, \bibinfo {author} {\bibfnamefont {D.~E.}\
  \bibnamefont {Reiter}}, \bibinfo {author} {\bibfnamefont {T.}~\bibnamefont
  {Kuhn}},\ and\ \bibinfo {author} {\bibfnamefont {R.~J.}\ \bibnamefont
  {Warburton}},\ }\bibfield  {title} {\bibinfo {title} {Demonstrating the
  decoupling regime of the electron-phonon interaction in a quantum dot using
  chirped optical excitation},\ }\href
  {https://doi.org/10.1103/PhysRevB.95.241306} {\bibfield  {journal} {\bibinfo
  {journal} {Phys. Rev. B}\ }\textbf {\bibinfo {volume} {95}},\ \bibinfo
  {pages} {241306(R)} (\bibinfo {year} {2017})}\BibitemShut {NoStop}%
\bibitem [{\citenamefont {Carmele}\ \emph {et~al.}(2010)\citenamefont
  {Carmele}, \citenamefont {Milde}, \citenamefont {Dachner}, \citenamefont
  {Harouni}, \citenamefont {Roknizadeh}, \citenamefont {Richter},\ and\
  \citenamefont {Knorr}}]{carmele2010formation}%
  \BibitemOpen
  \bibfield  {author} {\bibinfo {author} {\bibfnamefont {A.}~\bibnamefont
  {Carmele}}, \bibinfo {author} {\bibfnamefont {F.}~\bibnamefont {Milde}},
  \bibinfo {author} {\bibfnamefont {M.-R.}\ \bibnamefont {Dachner}}, \bibinfo
  {author} {\bibfnamefont {M.~B.}\ \bibnamefont {Harouni}}, \bibinfo {author}
  {\bibfnamefont {R.}~\bibnamefont {Roknizadeh}}, \bibinfo {author}
  {\bibfnamefont {M.}~\bibnamefont {Richter}},\ and\ \bibinfo {author}
  {\bibfnamefont {A.}~\bibnamefont {Knorr}},\ }\bibfield  {title} {\bibinfo
  {title} {Formation dynamics of an entangled photon pair: A
  temperature-dependent analysis},\ }\href
  {https://doi.org/10.1103/PhysRevB.81.195319} {\bibfield  {journal} {\bibinfo
  {journal} {Phys. Rev. B}\ }\textbf {\bibinfo {volume} {81}},\ \bibinfo
  {pages} {195319} (\bibinfo {year} {2010})}\BibitemShut {NoStop}%
\bibitem [{\citenamefont {Bracht}\ \emph {et~al.}(2022)\citenamefont {Bracht},
  \citenamefont {Seidelmann}, \citenamefont {Kuhn}, \citenamefont {Axt},\ and\
  \citenamefont {Reiter}}]{bracht2022phonon}%
  \BibitemOpen
  \bibfield  {author} {\bibinfo {author} {\bibfnamefont {T.~K.}\ \bibnamefont
  {Bracht}}, \bibinfo {author} {\bibfnamefont {T.}~\bibnamefont {Seidelmann}},
  \bibinfo {author} {\bibfnamefont {T.}~\bibnamefont {Kuhn}}, \bibinfo {author}
  {\bibfnamefont {V.~M.}\ \bibnamefont {Axt}},\ and\ \bibinfo {author}
  {\bibfnamefont {D.~E.}\ \bibnamefont {Reiter}},\ }\bibfield  {title}
  {\bibinfo {title} {Phonon wave packet emission during state preparation of a
  semiconductor quantum dot using different schemes},\ }\href
  {https://doi.org/https://doi.org/10.1002/pssb.202100649} {\bibfield
  {journal} {\bibinfo  {journal} {phys. status solidi (b)}\ }\textbf {\bibinfo
  {volume} {259}},\ \bibinfo {pages} {2100649} (\bibinfo {year}
  {2022})}\BibitemShut {NoStop}%
\bibitem [{\citenamefont {Wootters}(1998)}]{wootters1998entanglement}%
  \BibitemOpen
  \bibfield  {author} {\bibinfo {author} {\bibfnamefont {W.~K.}\ \bibnamefont
  {Wootters}},\ }\bibfield  {title} {\bibinfo {title} {Entanglement of
  formation of an arbitrary state of two qubits},\ }\href
  {https://doi.org/10.1103/PhysRevLett.80.2245} {\bibfield  {journal} {\bibinfo
   {journal} {Phys. Rev. Lett.}\ }\textbf {\bibinfo {volume} {80}},\ \bibinfo
  {pages} {2245} (\bibinfo {year} {1998})}\BibitemShut {NoStop}%
\bibitem [{\citenamefont {James}\ \emph {et~al.}(2001)\citenamefont {James},
  \citenamefont {Kwiat}, \citenamefont {Munro},\ and\ \citenamefont
  {White}}]{james2001measurment}%
  \BibitemOpen
  \bibfield  {author} {\bibinfo {author} {\bibfnamefont {D.~F.~V.}\
  \bibnamefont {James}}, \bibinfo {author} {\bibfnamefont {P.~G.}\ \bibnamefont
  {Kwiat}}, \bibinfo {author} {\bibfnamefont {W.~J.}\ \bibnamefont {Munro}},\
  and\ \bibinfo {author} {\bibfnamefont {A.~G.}\ \bibnamefont {White}},\
  }\bibfield  {title} {\bibinfo {title} {Measurement of qubits},\ }\href
  {https://doi.org/10.1103/PhysRevA.64.052312} {\bibfield  {journal} {\bibinfo
  {journal} {Phys. Rev. A}\ }\textbf {\bibinfo {volume} {64}},\ \bibinfo
  {pages} {052312} (\bibinfo {year} {2001})}\BibitemShut {NoStop}%
\bibitem [{\citenamefont {Bracht}\ \emph {et~al.}(2023)\citenamefont {Bracht},
  \citenamefont {Seidelmann}, \citenamefont {Karli}, \citenamefont {Kappe},
  \citenamefont {Remesh}, \citenamefont {Weihs}, \citenamefont {Axt},\ and\
  \citenamefont {Reiter}}]{bracht2023dressed}%
  \BibitemOpen
  \bibfield  {author} {\bibinfo {author} {\bibfnamefont {T.~K.}\ \bibnamefont
  {Bracht}}, \bibinfo {author} {\bibfnamefont {T.}~\bibnamefont {Seidelmann}},
  \bibinfo {author} {\bibfnamefont {Y.}~\bibnamefont {Karli}}, \bibinfo
  {author} {\bibfnamefont {F.}~\bibnamefont {Kappe}}, \bibinfo {author}
  {\bibfnamefont {V.}~\bibnamefont {Remesh}}, \bibinfo {author} {\bibfnamefont
  {G.}~\bibnamefont {Weihs}}, \bibinfo {author} {\bibfnamefont {V.~M.}\
  \bibnamefont {Axt}},\ and\ \bibinfo {author} {\bibfnamefont {D.~E.}\
  \bibnamefont {Reiter}},\ }\bibfield  {title} {\bibinfo {title} {Dressed-state
  analysis of two-color excitation schemes},\ }\href
  {https://doi.org/10.1103/PhysRevB.107.035425} {\bibfield  {journal} {\bibinfo
   {journal} {Phys. Rev. B}\ }\textbf {\bibinfo {volume} {107}},\ \bibinfo
  {pages} {035425} (\bibinfo {year} {2023})}\BibitemShut {NoStop}%
\bibitem [{\citenamefont {Schimpf}\ \emph {et~al.}(2021)\citenamefont
  {Schimpf}, \citenamefont {Reindl}, \citenamefont {Basso~Basset},
  \citenamefont {J{\"o}ns}, \citenamefont {Trotta},\ and\ \citenamefont
  {Rastelli}}]{schimpf2021quantumdots}%
  \BibitemOpen
  \bibfield  {author} {\bibinfo {author} {\bibfnamefont {C.}~\bibnamefont
  {Schimpf}}, \bibinfo {author} {\bibfnamefont {M.}~\bibnamefont {Reindl}},
  \bibinfo {author} {\bibfnamefont {F.}~\bibnamefont {Basso~Basset}}, \bibinfo
  {author} {\bibfnamefont {K.~D.}\ \bibnamefont {J{\"o}ns}}, \bibinfo {author}
  {\bibfnamefont {R.}~\bibnamefont {Trotta}},\ and\ \bibinfo {author}
  {\bibfnamefont {A.}~\bibnamefont {Rastelli}},\ }\bibfield  {title} {\bibinfo
  {title} {Quantum dots as potential sources of strongly entangled photons:
  Perspectives and challenges for applications in quantum networks},\ }\href
  {https://doi.org/10.1063/5.0038729} {\bibfield  {journal} {\bibinfo
  {journal} {Appl. Phys. Lett.}\ }\textbf {\bibinfo {volume} {118}},\ \bibinfo
  {pages} {100502} (\bibinfo {year} {2021})}\BibitemShut {NoStop}%
\bibitem [{\citenamefont {Hanschke}\ \emph {et~al.}(2018)\citenamefont
  {Hanschke}, \citenamefont {Fischer}, \citenamefont {Appel}, \citenamefont
  {Lukin}, \citenamefont {Wierzbowski}, \citenamefont {Sun}, \citenamefont
  {Trivedi}, \citenamefont {Vu{\v c}kovi{\'c}}, \citenamefont {Finley},\ and\
  \citenamefont {M{\"u}ller}}]{hanschke2018quantum}%
  \BibitemOpen
  \bibfield  {author} {\bibinfo {author} {\bibfnamefont {L.}~\bibnamefont
  {Hanschke}}, \bibinfo {author} {\bibfnamefont {K.~A.}\ \bibnamefont
  {Fischer}}, \bibinfo {author} {\bibfnamefont {S.}~\bibnamefont {Appel}},
  \bibinfo {author} {\bibfnamefont {D.}~\bibnamefont {Lukin}}, \bibinfo
  {author} {\bibfnamefont {J.}~\bibnamefont {Wierzbowski}}, \bibinfo {author}
  {\bibfnamefont {S.}~\bibnamefont {Sun}}, \bibinfo {author} {\bibfnamefont
  {R.}~\bibnamefont {Trivedi}}, \bibinfo {author} {\bibfnamefont
  {J.}~\bibnamefont {Vu{\v c}kovi{\'c}}}, \bibinfo {author} {\bibfnamefont
  {J.~J.}\ \bibnamefont {Finley}},\ and\ \bibinfo {author} {\bibfnamefont
  {K.}~\bibnamefont {M{\"u}ller}},\ }\bibfield  {title} {\bibinfo {title}
  {Quantum dot single-photon sources with ultra-low multi-photon probability},\
  }\href {https://doi.org/10.1038/s41534-018-0092-0} {\bibfield  {journal}
  {\bibinfo  {journal} {npj Quantum Information}\ }\textbf {\bibinfo {volume}
  {4}},\ \bibinfo {pages} {43} (\bibinfo {year} {2018})}\BibitemShut {NoStop}%
\end{thebibliography}%

\newpage
\appendix

\begin{table}[]
    \centering
    \caption{Parameters that are used in the simulations, unless mentioned otherwise.}
    \begin{tabular}{l l l}
    \hline\hline
    Parameter & Symbol & Value \\
    \hline
    Fine-structure splitting & $\hbar\delta_0$ & $\SI{0}{meV}$\\
     Biexciton binding energy & $\Delta_B$  & $\SI{1}{meV}$ \\
     Radiative decay rate, $X/Y$ & $\gamma_x$ & $\SI{0.01}{\per\pico\second}$\\
     Radiative decay rate, $B$ &  $\gamma_B = 2 \gamma_x$ \hspace{0.5cm} & $\SI{0.02}{\per\pico\second}$\\
     Cavity coupling & $\hbar g$ & $\SI{0.06}{meV}$\\
     Cavity outcoupling & $\hbar\kappa$ & $\SI{0.12}{meV}$\\
     Detuning, pulse 1 & $\Delta_1$ & $\SI{-5}{meV}$\\
     Detuning, pulse 2 & $\Delta_2$ & $\SI{-12.96}{meV}$\\
     Pulse area, pulse 1  & $\alpha_1$ & $32\pi$\\
     Pulse area, pulse 2 & $\alpha_2$ & $12.8\pi$\\
     Pulse durations & $\sigma_{1/2}$ & $\SI{2.7}{ps}$ \\
     \hline\hline
    \end{tabular}
    \label{tab:parameters}
\end{table}

\section{System Hamiltonian}
The quantum dot is modeled as a four-level system, which is placed inside a cavity with two cavity modes, one for each of the two orthogonal linear polarizations, $X$ and $Y$. The Hamiltonian of this system reads
\begin{align}
\begin{split}
    H_0 = &\hbar\omega_{x}(\ket{X}\bra{X} + \ket{Y}\bra{Y}) + (2\hbar\omega_{x} - \Delta_{B})\ket{B}\bra{B} \\&
    + (\hbar\omega_{x} - \Delta_{B}/2) (a^{\dagger}_X a^{\phantom{\dagger}}_X + a^{\dagger}_Y a^{\phantom{\dagger}}_Y),
\end{split}
\end{align}
where $\hbar\omega_x$ is the energy of the exciton states, $\Delta_B$ the biexciton binding energy (BBE) and the operators $a^{\phantom{\dagger}}_{X/Y} (a^{\dagger}_{X/Y})$ destroy (create) a photon in the respective cavity mode.
The quantum dot is excited via an external laser and coupled to the cavity, so the electron-light interaction is given by 
\begin{align}
\begin{split}
    H_{\text{el}} = &-\frac{\hbar}{2}(\Omega_X(t)\sigma^\dagger_X + \Omega_Y(t)\sigma^\dagger_Y)\\
    & + \hbar g \, (a_X\sigma^\dagger_X  +  \, a_Y\sigma^\dagger_Y) + \text{h.c.}.
\end{split}
\end{align}
Here, $g$ governs the strength of coupling to the cavity, and the terms $\Omega_{X/Y}(t)$ describe the field of the laser used to excite the quantum dot with the respective linear polarization operators $\sigma^{\phantom{\dagger}}_{S} = \ket{G}\bra{S} + \ket{S}\bra{B}$, where $S\in\{X,Y\}$. We assume a Gaussian-shaped pulse given by
\begin{equation}
    \Omega(t) = \frac{\alpha}{\sqrt{2\pi\sigma^2}}e^{-\frac{t^2}{2\sigma
^2}}e^{-i\omega_L t},
\end{equation}
where $\alpha$ is the pulse area, $\sigma$ is a measure of the pulse duration, which is related to the full-width at half maximum (FWHM) of the intensity by $\tau_{\text{FWHM}} = 2\sqrt{\ln(2)}\,\sigma$. The frequency of the laser pulse, denoted by $\omega_L$, is connected to the detuning to the quantum dot ground-state to exciton transition by $\Delta = \hbar(\omega_L - \omega_x)$.\\
The state preparation in quantum dots is disturbed by the surrounding environment. At low temperatures, the influence of longitudinal acoustic phonons acts as a limiting effect, this interaction with the lattice vibrations is modeled using the pure-dephasing type Hamiltonian
\begin{align}
\begin{split}
    H_\text{ph}= &\hbar \sum_{\mathbf{q}} \omega_{\mathbf{q}} b_{\mathbf{q}}^{\dagger} b_{\mathbf{q}}^{} +\hbar (\ket{X}\bra{X} + \ket{Y}\bra{Y} + \\
    & 2\ket{B}\bra{B} ) \sum_{\mathbf{q}} \left(g_{\mathbf{q}}^{} b_{\mathbf{q}}^{} + g_{\mathbf{q}}^* b_{\mathbf{q}}^{\dagger} \right)\,.
\end{split}
\end{align}
The operator $b_{\mathbf{q}}^{\dagger} (b^{}_\mathbf{q})$ creates (destroys) a phonon with wave vector $\mathbf{q}$. The phonons are coupled to the exciton states with the coupling element $g_\mathbf{q}$ and follow the linear dispersion relation $\omega_{\mathbf{q}} =c_{\text{LA}}q$, where $c_{\text{LA}}$ is the velocity of sound in the material. Containing two excitons, the coupling to the biexciton is twice as strong. We use the same material parameters as in Ref.~\cite{bracht2022phonon}, with an electron confinement length (size of the quantum dot) of $\SI{5}{nm}$.\\
The photons emitted by the quantum dot are modeled using Lindblad operators $\mathcal{L}$,
\begin{equation}
    \mathcal{L}_{{O},\gamma}\rho = \frac{\gamma}{2}\left(2{O}\rho{O}^\dagger - {O}^\dagger{O}\rho-\rho{O}^\dagger{O}\right).
\end{equation}
For the radiative decay with rate $\gamma$, this leads to the Operators $\mathcal{L}_{\ket{G}\bra{X},\gamma_x},\mathcal{L}_{\ket{G}\bra{Y},\gamma_x},\mathcal{L}_{\ket{X}\bra{B},\gamma_B/2},\mathcal{L}_{\ket{Y}\bra{B},\gamma_B/2}$. The out-coupling of photons from the cavity with rate $\kappa$ leads to the operators $\mathcal{L}_{a^{}_{X},\kappa}, \mathcal{L}_{a^{}_{Y},\kappa}$.

\section{Calculation of the Concurrence}
\begin{figure}
    \centering
    \includegraphics{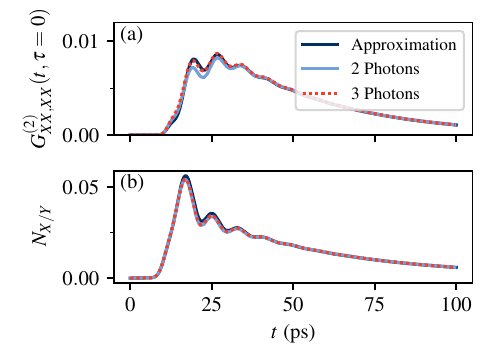}
    \caption{(a) Exemplary $G^{(2)}_{XX,XX}(t,\tau=0)$ for TPE, and (b) the dynamics of the cavity photon number $N_{X/Y}$ for calculations including two and three photons per cavity mode, compared to the approximation including one photon per cavity mode and the states $\ket{G,2,0}$ and $\ket{G,0,2}$.}
    \label{fig:approximation}
\end{figure}

To quantify the degree of entanglement, we utilize the concurrence as a measure of the entanglement degree \cite{wootters1998entanglement}.
The concurrence is determined from the two-photon density matrix $\rho^{\text{2P}}$ by evaluating the four eigenvalues $\lambda_i$ of the matrix
\begin{equation}
    M = \rho^{\text{2P}} T {\rho^{\text{2P}}}^* T ,
\end{equation}
where ${\rho^{\text{2P}}}^*$ represents the complex conjugate of the two-photon density matrix and $T$ is the anti-diagonal matrix with the elements $(-1,1,1,-1)$. After sorting the eigenvalues in decreasing order, i.e., $\lambda_{i+1}\le\lambda_i$, the concurrence is then given by \cite{james2001measurment,wootters1998entanglement}
\begin{equation}
    C = \text{max}\left\{0,\sqrt{\lambda_1}-\sqrt{\lambda_2}-\sqrt{\lambda_3}-\sqrt{\lambda_4}\right\}.
\end{equation}
The two-photon density matrix $\rho^{\text{2P}}$ is calculated using two-time correlation functions of the transition operators $\tilde{\sigma}_{X/Y}$, as explained in detail in Ref.~\cite{seidelmann2022two} (SI). In calculations without a cavity, the transition operators correspond to the polarization operators, i.e., $\tilde{\sigma}_{X/Y}=\sigma_{X/Y}$. In calculations including a cavity, the transition operators correspond to the cavity photon operators, i.e., $\tilde{\sigma}_{X/Y}=a^{\phantom{\dagger}}_{X/Y}$.  
These operators are then used in the two-time correlation functions of the form 
\begin{equation}
    G^{(2)}_{AB,CD}(t,\tau) = \langle \tilde{\sigma}_A^{\dagger}(t)\tilde{\sigma}_B^{\dagger}(t+\tau)\tilde{\sigma}_D(t+\tau)\tilde{\sigma}_C(t) \rangle.
\end{equation}
Due to numerical limitations, calculations including phonons and the cavity consider only one photon per $X/Y$ polarized cavity mode. In addition, the states consisting of the quantum dot's ground state and two photons per cavity mode (i.e., $\ket{G, n_X=2, n_Y=0}, \ket{G,n_X=0,n_Y=2}$) are included to ensure accurate results in the two-time correlation functions. Specifically, the correlation functions of the type $G^{(2)}_{XX,XX}(t,\tau=0) = \langle a^{\dagger}_X(t)a^{\dagger}_X(t)a^{\phantom{\dagger}}_X(t)a^{\phantom{\dagger}}_X(t)\rangle$ would always be zero if only one photon per cavity mode was considered. By using this approach, the Hilbert space dimension is reduced to 18 $(=4\times2\times2 +2)$ instead of 36 if two photons were fully included. This approximation significantly reduces the computation time, as the numerical effort including phonons scales unfavorably with the dimension. We have verified in the phonon-free case, comparing calculations made with the approximation and the complete inclusion of two and three photons per cavity mode, that including more photons has only negligible effects on the population dynamics and concurrence values for the parameter regime studied in this paper. Figure~\ref{fig:approximation} illustrates the impact of the approximation for the case of TPE (same parameters as in Fig.~\ref{fig:dynamics_compare_cavity}) without phonons. Panel (a) displays the two-time correlation $G^{(2)}_{XX,XX}(t,\tau=0)$, revealing only minor deviations around $t\sim\SI{20}{ps}$, when compared to calculations that fully include two or three photons per cavity mode. Similarly, panel (b) presents the dynamics of the cavity photon number, also demonstrating only slight deviations.

\section{Numerical Optimization for SUPER parameters}
\begin{figure}
    \centering
    \includegraphics{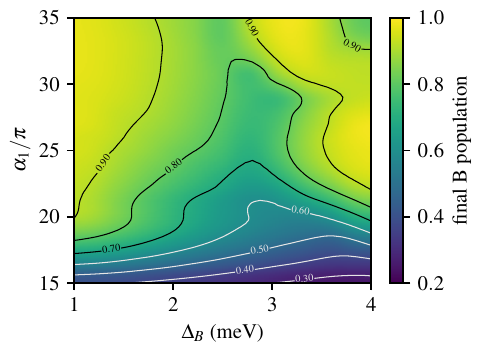}
    \caption{The final biexciton populations (disregarding decay, cavity coupling and phonons) for the same parameters as in Fig.~\ref{fig:area1_delta_b}: The parameters of the second pulse are optimized for a maximum final biexciton population.}
    \label{fig:b_final_values}
\end{figure}
In Fig~\ref{fig:area1_delta_b}, it was observed that a high concurrence exceeding $C=\SI{99}{\percent}$ could be achieved over a wide range of parameters. Here, the excitation parameters for each set were found through numerical optimization of the final biexciton occupation. Due to the computational complexity involved in calculating the concurrence values, the parameters for the SUPER scheme were optimized based on the biexciton population rather than directly optimizing the concurrence.
This approach is valuable in the way that it automatically provides results where a high photon rate can be expected. For each values of $\alpha_1$ and $\Delta_B$, with a fixed $\Delta_1=\SI{-5}{meV}$, the optimal $\alpha_2,\Delta_2$ were determined. The pulse areas $\alpha_{1/2}$ were constrained to a maximum value of $35\pi$.\\
Figure~\ref{fig:b_final_values} shows the final occupation of the biexciton state using the same parameters as in Fig~\ref{fig:area1_delta_b}. It is evident that for small pulse areas $\alpha_1$, the preparation fidelity drops rapidly. This outcome is expected, as previous studies demonstrated that a high pulse area has to be used for the scheme to work as intended \cite{bracht21swingup,bracht2023dressed}. Interestingly, for intermediate biexciton binding energies (BBEs), the preparation fidelity decreased to $\SI{70}{\percent}-\SI{80}{\percent}$. Due to the complex swing-up mechanism, there is no simple, straightforward explanation for this decrease in this parameter regime. We attribute it to the different system energies for varying BBEs that influence the mixing of the dressed states, leading to a more or less optimal preparation depending on the interplay of the dressed states as shown in Ref.~\cite{bracht2023dressed}. This finding highlights that the regimes for optimal concurrence and preparation fidelity (which, in turn, effects the photon rate) may differ. In the presented case, both reach high values for small biexciton bindings and high $\alpha_1$. The parameter set that yields the highest population, here for $\alpha_1=32\pi$, was then chosen for the further investigations.

\section{Impact of cavity coupling}
\begin{figure}
    \centering
    \includegraphics{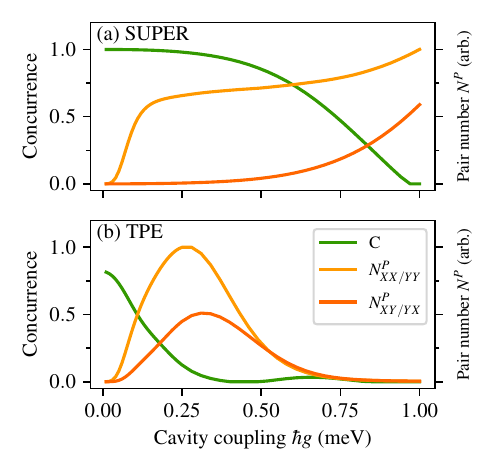}
    \caption{(a) Concurrence and the number of photon pairs emitted via the cavity for SUPER-excitation with the same parameters as in Fig.~\ref{fig:dynamics_compare_cavity}, depending on the dot-cavity coupling $g$. It is visible that for a strong dot-cavity coupling, the concurrence drops to zero. In this regime, the decay is strongly enhanced such that the quantum dot already emits the photons during the state preparation process, leading to photon pairs in the detrimental $\ket{XY}$ and $\ket{YX}$ state. (b) same for TPE, where due to the lack of the decoupling, the creation of detrimental photon pairs sets in immediately.}
    \label{fig:cavity_coupling}
\end{figure}
In all previous calculations involving a cavity, a constant cavity coupling strength of $\hbar g = \SI{0.06}{meV}$ was used alongside a constant cavity out-coupling rate of $\hbar\kappa = \SI{0.12}{meV}$.\\ 
Figure~\ref{fig:cavity_coupling} illustrates the influence of the cavity coupling on the concurrence and the number of photon pairs emitted via the cavity for (a) SUPER and (b) TPE. For SUPER, a plateau-like region emerges for cavity couplings up to approximately $\hbar g \sim \SI{0.2}{meV}$, beyond which the concurrence gradually decreases. The number of emitted photon pairs with the same polarization ($N^{P}_{XX/YY}$) rises sharply, as for small cavity couplings, the photons are emitted free-space before coupling to the cavity. With a smaller $\gamma$, as typically found in most quantum dots \cite{schimpf2021quantumdots,hanschke2018quantum}, an even greater share of photon pairs is emitted via the cavity, so that an arbitrarily high concurrence and a high photon yield can be achieved simultaneously.\\
As the cavity coupling strength increases, a larger portion of the emitted photons pass through the cavity. Eventually, almost the entire excitation of the quantum dot is transferred to cavity photon pairs. For very large couplings $\hbar g > \SI{0.75}{meV}$, additional photons are created due to re-excitation during the laser pulse.\\
The number of photon pairs with different polarizations ($N^{P}_{XY/YX}$) detrimental to the concurrence rises only slowly with increasing coupling values. This behavior can be largely contributed to the decoupling of the cavity from the QD during the preparation process. When the biexciton is prepared and no emission occurs during the pulses, photons are only emitted to the $\ket{XX}$ and $\ket{YY}$ states. A higher coupling efficiency increases the probability of photons being emitted already during the pulse.\\
In contrast, for TPE in a cavity, the absence of the decoupling mechanism results in a strong impact of the enhanced photon emission on the concurrence, as depicted in Fig.~\ref{fig:cavity_coupling}(b). Immediately, detrimental photon pairs are created during the preparation process, causing the concurrence to rapidly drop to zero. With increasing cavity couplings exceeding $\hbar g \sim \SI{0.3}{meV}$, the number of emitted photon pairs also decreases, as the strong energy shift resulting from the dot-cavity coupling hinders efficient population transfer.

\section{Influence of temperature on population dynamics}
\begin{figure}[thb]
    \centering
    \includegraphics{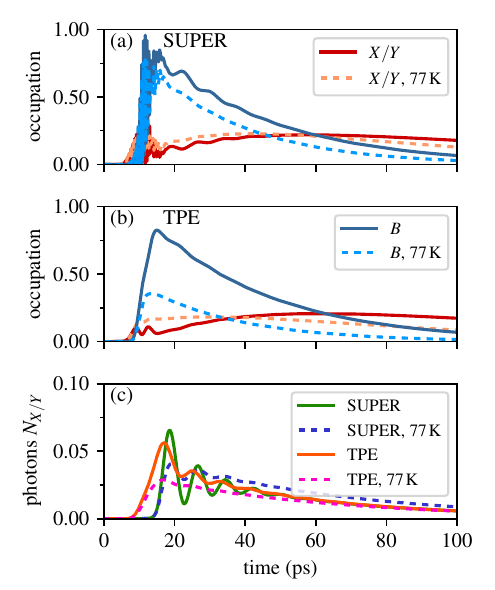}
    \caption{The influence of longitudinal acoustic phonons on the dynamics of (a) SUPER and (b) TPE in a cavity. The number of photons in the cavity is shown in (c).}
    \label{fig:cavity_phonons}
\end{figure}

In Figure~\ref{fig:temperature_dependence}, it was visible that the influence of the temperature shows significant differences between SUPER and TPE. Up to $\SI{77}{K}$, the final biexciton population only slightly decreases for SUPER, when compared to TPE. Involving a cavity, the concurrence basically remains a constant $\SI{99}{\percent}$ for SUPER, while it starts at about $\SI{69}{\percent}$ for TPE at $T=\SI{4}{K}$ and decreases with rising temperature. The findings for TPE are in agreement with previous studies \cite{heinze2017polarization,seidelmann2019phonon} that identified phonons as being a substantial source of decoherence, leading to a decrease of the concurrence. Phonons lead to renormalization of the cavity-dot coupling, effectively weakening the interaction \cite{seidelmann2019phonon}. The impact of phonons to the photon output can be seen in the population dynamics shown in Fig.~\ref{fig:cavity_phonons}. Panel (c) displays the number of $X/Y$ photons in the cavity, revealing that, when phonons are included, fewer photons are emitted into the cavity at early times compared to the phonon-free case. Additionally, the oscillations are damped. Panel (a) and (b) show the population dynamics of the dot states for SUPER and TPE, respectively, indicating that phonons disturb the process of TPE substantially more than SUPER.\\
\end{document}